\documentclass[letterpaper]{iopart}
\usepackage[fonts]{faust}
\usepackage{cite}
\usepackage{mathrsfs}
\usepackage{graphicx, amsmath, amssymb, amsbsy, latexsym, amsfonts, amsthm}
\usepackage[mathscr]{eucal}
\usepackage{datetime}

\DeclareMathOperator\svol{svol}

\newcommand\utilde[1]{#1}
\let\vec\acvec

\theoremstyle{plain}

\newcommand{\ident}{1}
\newcommand{\Cyl}{\mathrm{Cyl}}
\newcommand{\diff}{\mathrm{diff}}
\newcommand{\Diff}{\mathrm{Diff}}
\newcommand{\Cyldiff}{\Cyl_\diff^{\star}}
\newcommand{\CylDiff}{\Cyl_\Diff^{\star}}

\newcommand{\Hil}{\mathcal{H}}
\newcommand{\R}{\mathbb{R}}
\newcommand{\C}{\mathbb{C}}
\newcommand{\dual}{^\star}

\newcommand{\hol}{h}

\renewcommand{\Im}{\mathrm{Im}}
\renewcommand{\Re}{\mathrm{Re}}

\newcommand{\HIdec}{S}  % decoration indicating homogeneous isotropic model version
\newcommand{\Vsymm}{\mathcal{V}_{\mathrm{symm}}} % Symbol for quantum symmetric sector in full LQG,
\newcommand{\link}{\ell}

\newcommand{\newton}{\kappa} % 8\pi G

\newcommand{\matdec}[1]{#1}
\newcommand{\appref}[1]{\ref{#1}}

\begin{document}

\title{Diffeomorphism invariant cosmological sector in loop quantum gravity}
\author{C Beetle${}^1$, J S Engle${}^1$, M E Hogan${}^{1,2}$, and P Mendon\c{c}a${}^1$}
\address{${}^1$Department of Physics, Florida Atlantic University, 777 Glades Road, Boca Raton, FL 33431 USA}
\address{${}^2$Department of Mathematical and Physical Sciences, Wells College, 170 Main Street, Aurora, NY 13026, USA}
\ead{cbeetle@fau.edu, jonathan.engle@fau.edu, mhogan@wells.edu, pmendon1@fau.edu}

\bigskip
\begin{indented}
%\item \usdate \today
\item 7 June 2017
\end{indented}

\begin{abstract}

In this paper we work out in detail a new proposal to define rigorously a sector of loop quantum gravity at the diffeomorphism invariant level corresponding to homogeneous and isotropic cosmologies, and propose how to compare in detail the physics of this  sector with that of loop quantum cosmology. 
The key technical steps we have completed are
(a) to formulate conditions for homogeneity and isotropy in a diffeomorphism covariant way on the classical phase space of general relativity, and (b) to translate these conditions consistently using well-understood techniques to loop quantum gravity. 
To impose the symmetry at the quantum level, on both the connection and its conjugate momentum, 
the method used necessarily has similiarities to the Gupta-Bleuler method of quantizing the electromagnetic field.
Lastly, a strategy for embedding states of loop quantum cosmology into this new homogeneous isotropic sector, and using this 
embedding to compare the physics, is presented.

\end{abstract}

\maketitle

%\tableofcontents

\section{Introduction}

One of the most promising areas for extracting predictions from a theory of quantum gravity that can be compared meaningfully to observational data is in the application to cosmology.  Quantum effects are expected to dominate near the big bang, and different  models of quantum gravity make different predictions regarding the structure and dynamics of the gravitational field there.  However, the very early universe appears to have been homogeneous and isotropic to a remarkably high order of approximation, and almost  all models of the big bang itself treat those symmetries as exact.  This vastly simplifies the theoretical model, but raises fresh problems, particularly in the quantum context.  Specifically, homogeneity and isotropy constrain \textit{both} the configuration  and the momentum degrees of freedom, which is straightforward at the classical level, but is at odds with the Heisenberg uncertainty principle at the quantum level.  Accordingly, cosmological models impose symmetry at the classical level, prior to quantization.   Consequently, there is no direct connection between the quantum cosmological model used to extract predictions and the full theory of quantum gravity that one would like to constrain by comparing such predictions to observation.

This paper proposes a technique to establish such a direct connection between Loop Quantum Gravity (LQG) and Loop Quantum Cosmology (LQC).  The former is a promising framework for quantum general relativity, which has a well-understood kinematics and several  concrete proposals for its dynamics.  The latter is a quantization of the symmetric (\textit{i.e.}, homogeneous and isotropic) sector of classical general relativity whose kinematics is closely analogous to that of LQG.  The relative simplicity of LQC makes  possible a deeper and more complete analysis than has so far been possible for LQG.  In particular, one can solve the dynamics of LQC exactly, construct physical observables explicitly, and extract concrete predictions to compare with observation \cite{aps2006,  aps2006a, aps2006b, aan2012, aan2012a}.  However, it is important to remember that LQC \textit{is not derived} from LQG in any direct sense.  Rather, the two theories arise by applying similar mathematical quantization techniques to a pair of classical theories  related by symmetry reduction.  This leaves open the question of how insights from LQC can be used to help resolve ambiguities and constrain choices in LQG proper.

Our goal in this paper is to create a framework necessary to derive LQC from LQG, working at the quantum level throughout.  We do this by first identifying a sector of (distributional) quantum states in LQG that correspond to homogeneous and isotropic  spacetimes, and then examining how those states can be related to the states of LQC.  This direct approach, without a detour through classical physics, has a number of important benefits.

First, one must make certain choices in formulating LQC that have no analogues in the full theory.  In particular, there are at least two ambiguities that appear in LQC \textit{only} because one imposes symmetry before quantizing.  Recall that the standard  Hamiltonian constraint operator in LQG, due to Thiemann \cite{thiemann1996}, replaces the curvature appearing in the classical constraint with the holonomy around a loop, which in the end is shrunk to a point.  Thanks to diffeomorphism invariance, this limiting  process is independent of the details of the loop, whose area need not vanish in that limit because of the distributional nature of geometry in LQG.  In contrast, both diffeomorphism invariance and the distributional character of geometry are lost in LQC precisely  because homogeneity is imposed classically.  Consequently, the loop in the LQC Hamiltonian constraint cannot shrink to a point, and one must \textit{choose} its final area and shape.  One can motivate the conventional choice for the final area (see, \textit{e.g.},  the discussion at the end of \S III.B in \cite{aw2009}), but it is important to recognize that no analogous choices are necessary, or even possible, in the full theory.  A direct comparison with LQG at the quantum level can help shed light on this foundational  issue in LQC.

Second, a direct approach will enable positive developments on the LQC side to provide specific guidance in helping to specify the dynamics of LQG proper.  For example, it is currently unknown whether the leading proposal for the Hamiltonian constraint  operator $\hat{C}$ in LQG due to Thiemann \cite{thiemann1996} satisfies the correct Dirac constraint algebra.  The algebra can so far be tested only in a very trivial sense because the domain and co-domain of the constraint operator are mutually  exclusive.  It follows that the composition of two Hamiltonian constraints, and therefore their commutator, is not defined.  This lack of certainty regarding the Hamiltonian constraint makes it important to test the proposal in other ways, such as by relating  it to an LQC dynamics whose physical viability is known or can be more easily tested \cite{cs2008}.

The strategy developed here has some features in common with that described in a previous series of papers \cite{engle2006, engle2007, engle2008, engle2013}, but the two differ in several critical aspects.  Most importantly, while the previous papers \cite{engle2007,  engle2008, engle2013} also began by identifying a homogeneous and isotropic sector of states in LQG, they did so in a very different way.  Namely, they implemented the (classical) symmetry conditions via constraint functions that \textit{explicitly} break  diffeomorphism invariance because their definition relies on a specific, fixed action of the symmetry group.  After quantizing these symmetry conditions and imposing them as operator equations in LQG, the resulting symmetric sector therefore does not consist  of diffeomorphism-invariant states.  This poses a significant problem for the comparison with LQC, where the diffeomorphism constraint holds \textit{identically}.  Furthermore, the Thiemann Hamiltonian constraint in LQG is defined \textit{only} on diffeomorphism-invariant  states \cite{thiemann1996, rs1993}.  The homogeneous and isotropic sector defined in the previous papers therefore seems to have limited utility in relating the dynamics of LQG and LQC, which indeed is one of the central goals of the entire project.  This  latter problem could potentially be mitigated to a degree by group averaging the non-diffeomorphism invariant symmetric states \cite{engle2006, engle2007, engle2008, engle2012} in an appropriate sense.  But such a proposal would have to confront a large number  of ambiguities in making it concrete, which renders the approach quite unmanageable in practice.

In contrast, the present paper reformulates the definition of the homogeneous and isotropic sector in a way that preserves manifest diffeomorphism- (and gauge-)invariance throughout.  This permits a more faithful comparison of the symmetric sector of LQG  with the space of LQC states.  It also makes possible a detailed comparison of the dynamics of the two theories without gauge fixing.  As in the previous series of papers \cite{engle2007, engle2008, engle2013}, the basis for this comparison will be an embedding  map $\iota : \mathcal{H}_{\text{LQC}} \hookrightarrow \mathcal{V}_{\text{HI}}$ from the Hilbert space of LQC to the homogeneous and isotropic sector of LQG, which in the present case is a subspace of the space $\text{Cyl}^*_{\text{diff}}$ of diffeomorphism-invariant  LQG states.  The relative simplicity of the framework presented here (\textit{i.e.}, the absence of group averaging) will allow us to consider the criteria to be imposed on such an embedding in significantly more detail.

A companion paper \cite{behm2017} to this one will present a proof of concept for the framework developed here by applying the general ideas to the symmetry reduction from Bianchi I LQC to fully homogeneous and isotropic LQC.  (The natural embedding $\iota$  in that reduction turns out to be closely related to the projection mapping proposed in \cite{aw2009}, which also examines the relationship between these two cosmological models.)  The key results of this paper and its companion are also summarized, albeit  without proof, in \cite{behm2016}.

The remainder of this paper is organized as follows.  Section 2 establishes notation and conventions for LQG and LQC.  Section 3 describes the diffeomorphism-invariant constraint functions that we use to impose homogeneity and isotropy in classical general  relativity in a way that will translate straightforwardly to the loop quantization.  A key result of this section, and indeed of this paper, concerns the Poisson algebra of these constraint functions, together with the usual Gauss and diffeomorphism constraints  of general relativity, which we show closes in a way that will allow them to be imposed consistently (barring anomalies) at the quantum level.  Section 4 presents the quantum version of the symmetry conditions.  Section 5 analyzes properties and proposals  for the embedding mapping from the LQC Hilbert space into the symmetric sector of LQG.  Section 6 examines some detailed proposals for such an embedding, which will be explored further in subsequent work.  Section 7 concludes with a brief discussion.  Finally,  two appendices collect some technical details regarding the symmetry conditions and a toy model that helps to clarify the quantum embedding strategy.

\section{Preliminaries}

%%%%%%%%%%%%%%%%%%%%%%%%%%%%%%% Begin 2.1 %%%%%%%%%%%%%%%%%%%%%%%%%%%%%%%%%%%%%%%%%%%%%%%%

The basic variables of the classical theory underlying loop quantum gravity are an ${\rm SU(2)}$ connection $A^i_a$, called the Ashtekar--Barbero connection \cite{barbero1995}, and a spatial triad $\tilde{E}^a_i := \abs{\det e}\, e^a_i$ of density weight  $+1$ on the space-like hypersurface $M$ of the canonical theory.  Here, $a$ is a tangent-space index and $i$ is an index in an internal space that can be identified with the Lie algebra of SU(2).  Meanwhile, $e^a_i$ is an ordinary triad on $M$ that induces  a Riemannian metric $q_{ab} := e^i_a\, e_{bi}$.  For the purpose of this paper, in order to avoid spurious infinities, we take $M$ isomorphic to the 3-torus $T^3$.  (This is the full theory analogue of the `fiducial cell' usually used in introducing LQC \cite{aps2006}.)   In terms of the standard generalized ADM variables \cite{adm1962}, the Ashtekar--Barbero connection is 
\begin{equation}
A^i_a \equiv \Gamma^i_a + \gamma K^i_a ,
\end{equation}
where $\Gamma^i_a$ is the spin connection determined by $\tilde{E}^a_i$, and $K^i_a := K_{ab}\, e^{bi}$ with $K_{ab}$ the extrinsic curvature of $M$.  The Barbero--Immirzi parameter $\gamma \in \R^+$ \cite{barbero1995, immirzi1997} is a real constant that  can be fixed through considerations of black hole entropy \cite{meissner2004, abbdv2009, enpp2011, abk2000, enpp2010}.  The Poisson brackets of the basic variables in loop gravity are  
\begin{equation}
\pbrack[\big]{A^i_a(x)}{\tilde{E}^b_j(y)} 
= \newton \gamma\, \delta^b_a\, \delta^i_j\, \delta_x(y), 
\end{equation}
where $\newton := 8\pi G_{\text{Newton}}$.

The basic variables of loop gravity that have direct quantum analogues are \emph{holonomies} $A[\link]$ of the connection $A$ along curves $\link$, and the ``electric'' fluxes 
\begin{equation}
\Sigma[S,f] := \int_S \Sigma^i\, f_i
\end{equation}
through arbitrary 2-surfaces $S$, where $\Sigma^i_{ab} := \utilde{\eta}_{abc}\, \tilde{E}^{ci}$ with $\utilde{\eta}_{abc}$ the Levi-Civita density of weight $-1$ on $M$, and $f_i$ is an arbitrary smearing field \cite{ashtekar1991}.

States of LQG in the \emph{connection representation} are functionals $\Psi(A)$ of the connection.  One starts with a space, denoted $\Cyl$, of ``nicely-behaved'' \textit{cylindrical} functions, which depend on the connection $A$ \emph{only} through its  holonomies $A[\ell]$ along a finite set of piecewise-analytic curves $\ell$. The natural Ashtekar--Lewandowski inner product $\iprod{\cdot}{\cdot}$ is induced on $\Cyl$ essentially from the Haar measure on SU(2) \cite{ai1992, al1994}. The elementary quantum  operators $\hat{A}[\link]$ and $\hat{\Sigma}[S,f]$ are naturally defined on a kinematical Hilbert space $\Hil_{\mathrm{kin}}$, which can be obtained by completing $\Cyl$ in this inner product.  This kinematical Hilbert space has a natural orthonormal basis  associated with (generalized) \emph{spin-network states} $|\gamma, \vec{j}, \vec{T} \rangle \in \Cyl$.  Each such state is labelled by a graph $\gamma$ consisting of a finite set of curves, called edges, intersecting at most at their end points, called nodes.   Each edge is decorated with a ``spin'' label $j$ that specifies an irreducible representation of SU(2), and each node is decorated with a tensor $T$ in the product of the representations on the adjacent edges \cite{al2004, rovelli2004, thiemann2007}.  The  spin-network states are eigenstates of the operators corresponding to the areas of 2-surcfaces $S \subset M$, whose spectra are discrete and determined entirely by the spin labels $j$.

Many physically interesting states of LQG, including both diffeomorphism-invariant states and states in the homogeneous and isotropic sector of LQG defined below, live in the algebraic dual $\Cyl^\star$ of $\Cyl$. We refer to these generally as \emph{distributional}  states. To emphasize their distributional nature, elements $\bra[][(|]{\Psi} \in \Cyl^\star$ are denoted using a ``rounded'' bra \cite{afw2002}.  The Ashtekar--Lewandowski inner product $\iprod{\cdot}{\cdot}$ defines a natural  embedding of $\Cyl \subset \mathcal{H}_{\text{kin}}$  into $\Cyl^{\star}$.  It is preferable, for a number of reasons, that this embedding be a linear map, whence we choose the natural \textit{physical} vector structure of $\Cyl^{\star}$ such that 
\begin{equation}
\braket[\big][(>]{\Psi_1 + \lambda \Psi_2}{\Phi} 
:= \braket[\big][(>]{\Psi_1}{\Phi} + \bar\lambda\, \braket[\big][(>]{\Psi_2}{\Phi} 
\end{equation}
for all $\ket{\Phi} \in \Cyl \subset \mathcal{H}_{\text{kin}}$.  Diffeomorphisms act naturally on $\Cyl$, and hence also on $\Cyl^\star$. The subspace of $\Cyl^\star$ that is invariant under this action --- \textit{i.e.}, the space of solutions to the  diffeomorphism constraint --- is denoted $\CylDiff$. A group-averaging map $\eta: \Cyl \rightarrow \Cyl^\star_\Diff$  \cite{almmt1995} induces a natural inner product $\iprod{\cdot}{\cdot}_\diff$ on the image $\Cyl_\diff \subset \Cyl^\star_\Diff$ of $\eta$  via $\langle \eta \Phi_1, \eta \Phi_2 \rangle := (\eta \Phi_1|\Phi_2 \rangle$.  The completion of $\Cyl_\diff$ in this inner product defines the diffeomorphism-invariant Hilbert space $\Hil_\diff$ of LQG.

The algebraic dual $\Cyl_\diff^\star$ of $\Cyl_\diff$ is naturally embedded into $\Cyl_\Diff^\star$ via the map 
\begin{equation}
I : \Cyl_\diff^\star  \rightarrow \Cyl_\Diff^\star 
\qquad\text{with}\qquad 
\braket[\big][(>]{I\Psi}{\phi} := \Psi (\eta \phi), 
\end{equation}
where $\Psi \in \Cyl_\diff^\star$ and $\phi \in \Cyl$ are arbitrary.  The diffeomorphism invariance of $I\Psi$ is immediate from this definition, whence it indeed is an element of $\Cyl_\Diff^\star$.  The embedding map $I$ is also easily checked to be  injective.  The standard injection $\Hil_\diff \hookrightarrow \Cyl_\diff^\star$, which sends $\Psi \mapsto \iprod{\Psi}{\cdot}_\diff$, together with $I$, therefore provides an embedding of $\Hil_\diff$ into $\Cyl_\diff^{\star}$.  It follows that $\Cyl_\Diff^\star$  is a natural, ``universal home'' for all diffeomorphism invariant states, whether normalizable with respect to $\iprod{\cdot}{\cdot}_\diff$ or not.

With the basic architecture of the full theory established, we turn to the simplified theory of \emph{loop quantum cosmology} (LQC) \cite{as2016, ac2013, as2011a, bcm2011, aps2006, abl2003}.  Consider the spatially flat ($k=0$) case, where the relevant  symmetry is the three-dimensional Euclidean group $\mathcal{E}$.  LQC begins by \emph{fixing} an action of $\mathcal{E}$ on $M$.  Let $\mathring e^a_i$ denote a reference triad on $M$ that is invariant under the action of $\mathcal{E}$.  The general state  $(A, \tilde E)$ of classical general relativity that is invariant under the same action of $\mathcal{E}$ has the form 
\begin{equation}\label{initcrit}
A_a^i = c\, \mathring e_a^i 
\qquad\text{and}\qquad 
\tilde E^a_i = p\, \abs{\det\mathring e}\, \mathring e^a_i, 
\end{equation}
where $c$ and $p$ are both constant over $M$.  These constants are the basic variables of LQC.  The symplectic structure for this cosmological model arises by restricting the integral over $M$ from the full theory to a finite spatial region called the  \emph{fiducial cell}.  The basic Poisson bracket in this case is 
\begin{equation}
\pbrack{c}{p} = \frac{\newton\gamma}{3 \mathring{V}}, 
\end{equation}
where $\mathring{V}$ is the volume of the fiducial cell in the reference geometry $\mathring e^a_i$.

As in the full theory, there is no operator corresponding to the reduced connection $c$ in LQC, but there are operators corresponding to its holonomies (along the straight lines dictated by the fixed action of $\mathcal{E}$).  These holonomies can be expressed  in terms of the operator analogues $\widehat{\vphantom{t}\ee^{\smash{\ii\mu c}}}$ of the exponentials of $c$.  States in the connection representation of LQC belong to a certain class of functions $\psi(c)$.  To specify this class, we recall that $\psi(c)$  is \emph{almost periodic} if it is a (finite) linear combination of exponentials $\ee^{\ii\mu c}$ \cite{abl2003}.  The space of such functions is analogous to the space of cylindrical functions in the full theory, whence it is denoted $\Cyl_S$.  There is a  natural inner product $\iprod{\cdot}{\cdot}$ on the space of almost periodic functions such that the basic operators $\widehat{\vphantom{t}\ee^{\smash{\ii\mu c}}}$ and $\hat p$ satisfy appropriate reality conditions.  Completing $\Cyl_S$ in this inner product  yields the (kinematical) Hilbert space $\mathcal{H}_S$ of LQC.  We will denote states in $\mathcal{H}_S$ using the usual bra-ket notations, $\ket{\psi}$ and $\bra{\phi}$.  The operator $\hat p$ on $\mathcal{H}_S$ analogous to the flux operators in the full  theory has a \emph{discrete} spectrum in the sense that it has a complete basis of \emph{normalizable} eigenstates $\ket{p}$.  (This basis is uncountable, however, as the eigenvalue $p$ can take arbitrary real values.)  The volume of the fiducial cell corresponds  to the operator $\hat V := \mathring{V}\, \abs{\hat p}{}^{3/2}$, and so can be diagonalized in the same discrete basis.  The kinematics described here for the $k=0$ case has been recently shown to be uniquely determined by invariance under residual diffeomorphisms  \cite{eh2016, eht2016}.

One obtains a similar quantum kinematics in the $k=1$ \cite{apsv2006, skl2006} and $k=-1$ \cite{vandersloot2006} cases.  Once again, the system is described by a single pair of phase-space degrees of freedom $(c,p)$, and one can define corresponding operators  $\widehat{\vphantom{t}\ee^{\smash{\ii\mu c}}}$ and $\hat{p}$ in the quantum theory.  The volume of the fiducial cell is given by $\hat{V} = \mathring{V}\, \abs{\hat{p}}{}^{3/2}$, and the space of ``nice'' states on which these operators are well-defined is  the space of almost periodic functions $\Cyl_S$.  The eigenstates of $\hat{p}$, and hence of $\hat{V}$, remain normalizable in the natural inner product.

\section{The symmetry constraints: Classical phase space analysis}
\label{sec:classical}

\newcommand\bip{\gamma}
\newcommand\cs{\alpha}
\newcommand\vq{v_0}
\newcommand\ED{{\mathrm{D}}}
\newcommand\rKop{\mathrm{K}}
\newcommand\cKop{\mbb{K}}

This section formulates conditions that select those points in the phase space of classical general relativity corresponding to homogeneous and isotropic cosmologies.  These conditions must restrict the spatial geometry at a given moment of time to be  maximally symmetric, as well as the canonical momenta such that the spatial geometry remains maximally symmetric under time evolution.  Our ultimate goal,
which we will complete in the next section,
is to promote these symmetry conditions to the full quantum theory in terms of operators defined on the Hilbert space of loop quantum gravity.  There are several key points, both technical and conceptual, to be addressed. 
 
Technically, we must formulate the classical symmetry conditions purely in terms of functions on the phase space of general relativity that can be promoted to specific and well-defined operators in the quantum theory.  In order to preserve diffeomorphism  invariance, these functions cannot refer to a fixed action of a particular symmetry group.  (Recall that, in contrast, the usual ``mini-superspace'' approach to quantum cosmology begins precisely by fixing such a group action, thereby explicitly breaking the  diffeomorphism gauge symmetry at the classical level, prior to quantization.)  Accordingly, in subsection \ref{SC.covar} we describe a set of \textit{covariant} conditions that identify homogeneous and isotropic sets of Cauchy data within the full phase space  of general relativity while leaving diffeomorphism invariance intact.  In subsection \ref{SC.quant} we cast those symmetry conditions as the vanishing $\mbb{S}[f, g] = 0$ of a family of (complex-valued) phase-space functions, parameterized by a pair $(f_{ij},  g_{kl})$ of smearing fields.  We also sketch how to promote those functions to concrete operators on Hilbert space in the quantum theory.
 
The main conceptual challenge is that the classical symmetry conditions necessarily constrain both configuration and momentum degrees of freedom simultaneously.  While this is straightforward classically, some care is needed at the quantum level because  it is generally impossible to impose such simultaneous constraints precisely (\textit{i.e.}, as operator equations).  However, we show in subsection \ref{SC.palg} that the classical Poisson algebra of the complex-valued functions $\mbb{S}[f, g]$ from subsection  \ref{SC.quant} \textit{closes}.  (This is no longer the case if one extends that algebra to include the complex conjugate functions $\bar{\mbb{S}}[f, g]$ as well.)  It follows that, barring anomalies arising from quantization, the conditions $\hat{\mbb{S}}[f,  g] \ket{\psi} = 0$ are mutually consistent at the quantum level, and thus can be imposed simultaneously.  This will define the symmetric sector of loop quantum gravity.  A similar symmetry reduction scheme exists for scalar field models \cite{engle2006}, where  the non-Hermitian operators analogous to $\hat{\mbb{S}}[f, g]$ are annihilation operators on Fock space and the conditions analogous to $\hat{\mbb{S}}[f, g] \ket\psi = 0$ assert that all non-symmetric field modes are \textit{unexcited}.  (The Gupta--Bleuler  quantization \cite{gupta1950, bleuler1950} of the electromagnetic field rests on essentially the same idea.)  Importantly, this quantum symmetry reduction of scalar fields commutes with quantization in a precise sense \cite{engle2006}.  There is no comparably  straightforward interpretation of the quantum symmetry conditions in the gravitational case, and we will see in later sections that there are additional technical subtleties to confront at the quantum level as well.  Nonetheless, the closure of the Poisson  algebra of classical symmetry conditions $\mbb{S}[f, g]$ established in subsection \ref{SC.palg} is critically important to the definition of the homogeneous isotropic sector in full loop quantum gravity we propose.
 
Rounding out the classical analysis, in subsection \ref{SC.dynam} we examine the interplay of the symmetry conditions $\mbb{S}[f, g] = 0$ and the constraints of (Euclidean) general relativity.  The main result here is that the extended Poisson algebra  including the symmetry conditions $\mbb{S}[f, g]$ and the Euclidean Hamiltonian constraint $C_{\mathrm{E}}[N]$ does \textit{not} close.  The bracket $\pbrack{\mbb{S}[f, g]}{C_{\mathrm{E}}[N]}$ vanishes classically, provided the symmetry conditions hold and  the lapse $N$ is chosen to be uniform in space.  But it contains terms proportional to the complex conjugate symmetry conditions $\bar{\mbb{S}}[h, k]$.  It follows that we should \textit{not} expect the Hamiltonian constraint operator to preserve the quantum  symmetric sector.  Rather, we will have to use weaker notions, based for example on expectation values or matrix elements, to characterize the dynamics induced by the full theory on its symmetric sector.

\subsection{Diffeomorphism covariant symmetry conditions}
\label{SC.covar}
 
The spacetime of a homogeneous and isotropic cosmology is foliated by spacelike hypersurfaces $\Sigma$ such that the intrinsic metric $g_{ab}$ on each admits the maximum number $d (d + 1) / 2$ of Killing fields, where $d$ is the dimension of $\Sigma$.   The curvature tensor of a maximally symmetric metric $g_{ab}$ satisfies
\begin{equation}\label{SC:sRiem}
                R_{abcd} = \frac{2 R}{d (d - 1)}\, g_{c[a}\, g_{b]d},
\end{equation}
where the scalar curvature $R$ is constant throughout $\Sigma$.  Conversely, \ref{app:maxsym} shows that if the curvature of a Riemannian geometry $(\Sigma, g_{ab})$ satisfies (\ref{SC:sRiem}) for some constant $R$, then $g_{ab}$ necessarily admits the  maximal number of Killing fields throughout $\Sigma$.  Thus, the local and diffeomorphism-covariant relation (\ref{SC:sRiem}) between geometric fields is equivalent to the global symmetry of the metric ordinarily imposed by demanding invariance under a fixed  action of a symmetry group on $\Sigma$.  (Note that this condition implies only that there is \textit{a} maximal symmetry group, whose structure is determined by the sign of $R$.)  Similarly, the condition for (\ref{SC:sRiem}) to continue holding under time  evolution  can be written in the diffeomorphism-covariant form
\begin{equation}\label{SC:sExtC}
                K_{ab} = H\, g_{ab},
\end{equation}
where again $H$ is constant throughout $\Sigma$.  These conditions translate immediately into the relations
\begin{equation}\label{SC:sAshD}
                {}^\Gamma\! F_{ab}{}^i = \rho\, \Sigma_{ab}{}^i
                \qquad\text{and}\qquad
                K_a^i := \frac{A_a^i - \Gamma_a^i}{\bip} = H\, e_a^i
\end{equation}
between the basic triad and connection fields used in the loop quantization of general relativity, where $\Gamma_a^i$ denotes the spin connection of $e_a^i$, ${}^\Gamma\! F_{ab}{}^i$ denotes its curvature, and $\Sigma_{ab}{}^i := \epsilon^i{}_{jk}\, e^j_a\,  e^k_b$.
 
The two classical symmetry conditions (\ref{SC:sAshD}) can be combined further into a single condition in terms of a complexified Ashtekar connection.  This complexified connection is defined by
\begin{equation}\label{SC:comA}
                \mbb{A}_a^i := A_a^i + \ii \cs\, e_a^i,
\end{equation}
where $\cs$ is an arbitrary, real constant with units of inverse length.  The curvature of $\mbb{A}_a^i$ is
\begin{align}\label{SC:comF}
                \mbb{F}_{ab}{}^i
                                := \ed \mbb{A}_{ab}{}^i + \epsilon^i{}_{jk}\, \mbb{A}_a^j\, \mbb{A}_b^k
                                &= F_{ab}{}^i + \ii \cs\, D e_{ab}{}^i - \cs^2\, \Sigma_{ab}{}^i
                                                \notag\\[1ex]
                                &= {}^\Gamma\! F_{ab}{}^i + \bip\, {}^\Gamma\! D K_{ab}{}^i
                                                + \bip^2\, \epsilon^i{}_{jk}\, K_a^j\, K_b^k - \cs^2\, \Sigma_{ab}{}^i
                                                + 2 \ii \cs \bip\, \epsilon^i{}_{jk}\, K_{[a}{}^j\, e_{b]}{}^k,
\end{align}
where $D$ and ${}^\Gamma\! D$ denote the covariant exterior derivatives associated with the (real) Ashtekar and spin connections, respectively.  If the symmetry conditions (\ref{SC:sAshD}) hold, then
\begin{equation}\label{SC:sComD}
                \mbb{F}_{ab}{}^i
                                = \mbb{b}\, \Sigma_{ab}{}^i
                                := \bigl( \rho + \bip^2 H^2 - \cs^2 + 2 \ii \cs \bip H \bigr)\, \Sigma_{ab}{}^i,
\end{equation}
where $\mbb{b}$ is complex and constant throughout $\Sigma$.  Conversely, if (\ref{SC:sComD}) holds, then transvecting its imaginary part with $\epsilon_i{}^{lm}\, e^b_m$ gives
\begin{equation}\label{SC:sComK}
                2 \imag\mbb{b}\, e^l_a = \cs\bip\, K_a^l + \cs \bip\, K_b^m\, e_m^b\, e^l_a
                \qquad\leadsto\qquad
                K_a^l = \frac{\imag\mbb{b}}{2 \cs \bip}\, e_a^l.
\end{equation}
The result on the right follows by further transvecting the result on the left with $e_l^a$ to evaluate its last term.  Substituting this extrinsic curvature into the real part of the complex curvature in (\ref{SC:comF}) then gives
\begin{equation}\label{SC:sComF}
                {}^\Gamma\! F_{ab}{}^i
                                = \biggl(\real\mbb{b} - \frac{(\imag \mbb{b})^2}{4 \cs^2} + \cs^2 \biggr)\, \Sigma_{ab}{}^i.
\end{equation}
Thus, the real and imaginary parts of the single complex relation (\ref{SC:sComD}) imply both of the relations (\ref{SC:sAshD}) that characterize homogeneous and isotropic Cauchy data.

\subsection{Quantizing the symmetry conditions}
\label{SC.quant}
 
Now we turn to the question of how the proportionality (\ref{SC:sComD}) between $\mbb{F}_{ab}{}^i$ and $\Sigma_{ab}{}^i$ can be implemented in terms of operators on the Hilbert space of loop quantum gravity.  The key idea is to use Thiemann's complexifier  technique \cite{thiemann2002} to define quantum operators corresponding to holonomies of the complexified connection (\ref{SC:comA}).  Let us first recall the complexifier technique in general.
 
Fix a (real-valued) function $C$ on phase space, the \textit{complexifier}.  Then, given any other (real-valued) observable $O$, define a 1-parameter family of complex-valued observables ${}^t \mbb{O}$ by setting
\begin{equation}\label{SC:ccFlow}
                \pdby{t}\, {}^t \mbb{O} := \ii\, \pbrack[\big]{{}^t \mbb{O}}{C}
                \qquad\text{with}\qquad
                {}^0 \mbb{O} := O.
\end{equation}
The complexification of $O$ is $\mbb{C}(O) := \mbb{O} := {}^1 \mbb{O}$.  We refer to (\ref{SC:ccFlow}) as the \textit{complexification flow}.  If $O$ and $C$ have well-defined quantum analogues $\hat O$ and $\hat C$, respectively, then the corresponding  quantum flow is
\begin{equation}\label{SC:qcFlow}
                \pdby{t}\, {}^t \hat{\mbb{O}} := \frac{1}{\hbar}\, \comm[\big]{{}^t \hat{\mbb{O}}}{\hat C}
                \qquad\text{with}\qquad
                {}^0 \hat{\mbb{O}} := \hat O.
\end{equation}
It is straightforward to integrate these equations in the quantum case to find
\begin{equation}\label{SC:qcOp}
                {}^t \hat{\mbb{O}} = \ee^{- t \hat C / \hbar}\, \hat O\, \ee^{t \hat C / \hbar}
\end{equation}
This solution of the flow equations (\ref{SC:qcFlow}) makes a number of important results immediate at the quantum level.  In particular, it is clear that the complexification $\mbb{C}(\hat O_1 \hat O_2)$ of a product of operators is equal to the product  $\mbb{C}(\hat O_1)\, \mbb{C}(\hat O_2)$ of their separate complexifications, and thus that the complexification $\mbb{C} \bigl( \comm[\nml]{\hat O_1}{\hat O_2} \bigr)$ of a commutator equals the commutator $\comm[\big]{\mbb{C}(\hat O_1)}{\mbb{C}(\hat O_2)}$  of the separate complexifications.  The corresponding results also hold at the classical level.
That is,
\begin{align*}
\mbb{C}(O_1 O_2) = \mbb{C}(O_1) \mbb{C}(O_2)
\quad \text{and} \quad 
\{\mbb{C}(O_1), \mbb{C}(O_2)\} = \mbb{C}(\{O_1, O_2\}) .
\end{align*}
These follow from the Leibniz property
\begin{equation}\label{SC:pbLeib}
                \pbrack{f g}{C} = \pbrack{f}{C}\, g + f\, \pbrack{g}{C}
\end{equation}
of the Poisson bracket and the Jacobi identity
\begin{equation}
                \pbrack[\big]{\pbrack{f}{g}}{C} = \pbrack[\big]{\pbrack{f}{C}}{g} + \pbrack[\big]{f}{\pbrack{g}{C}},
\end{equation}
respectively.
 
Now we focus on the specific application of the complexifier technique used here.  Choosing $C$ to be proportional to the total volume $V$ of space leads to the complexified connection (\ref{SC:comA}).  Recall that
\begin{equation}
                \pbrack[\big]{A_a^i(x)}{V}
                                = \int_y \frac{1}{2}\, \abs[\big]{\det E(y)}^{1/2}\, E_b^j(y)\, \pbrack[\big]{A_a^i(x)}{E^b_j(y)}
                                = \frac{\newton\bip}{2}\, e_a^i(x)
\end{equation}
and $\pbrack[\big]{e_a^i(x)}{V} = 0$.  It follows that
\begin{equation}\label{SC:Aflow}
                \pdby{t}\, {}^t\! \mbb{A}_a^i
                                = \ii\, \pbrack{{}^t\! \mbb{A}_a^i}{\frac{2 \alpha}{\newton \bip}\, V}
                                = \ii \alpha\, e_a^i
                \qquad\leadsto\qquad
                {}^t\! \mbb{A}_a^i = {}^0\! \mbb{A}_a^i + \ii \alpha t\, e_a^i.
\end{equation}
Setting ${}^0\! \mbb{A}_a^i := A_a^i$ yields the complexified connection (\ref{SC:comA}) for $t = 1$.
 
There is no operator corresponding to the classical connection $A_a^i$ in loop quantum gravity, but there are operators corresponding to the holonomies $A[\link] := \hol[A, \ell]$ of that connection along (piecewise analytic) curves $\link \subset \Sigma$.   But it is straightforward to show that the complexifier flow
\begin{equation}\label{SC:hflow}
                \pdby{t}\, {}^t \mbb{A}[\link] = \ii\, \pbrack{{}^t \mbb{A}[\link]}{\frac{2 \alpha}{\newton \bip}\, V}
                \qquad\text{with}\qquad
                {}^0 \mbb{A}[\link] := A[\link]
\end{equation}
of a given holonomy $A[\link]$ gives precisely the holonomy $\hol[{}^t\! \mbb{A}, \ell]$ of the complexified connection from (\ref{SC:Aflow}).  It follows that it is equally straightforward to define quantum operators
\begin{equation}\label{SC:chop}
                \hat{\mbb{A}}[\ell] := \ee^{- \hat V / v_0}\, \hat A[\ell]\, \ee^{\hat V / v_0}
\end{equation}
corresponding to holonomies of the complexified connection $\mbb{A}_a^i$.  Here, $v_0 := \hbar \newton \bip / 2 \alpha$ is an arbitrary constant with units of volume since $\alpha$ was arbitrary with units of inverse length.  These operators are, at least  in principle, well-defined on the Hilbert space of loop quantum gravity.
 
The challenge now is to express the tensorial symmetry condition (\ref{SC:sComD}) in terms of the complexified holonomies (\ref{SC:hflow}).  One can do this using existing techniques.  Form the wedge product of either side of (\ref{SC:sComD}) with $\sgn(\det  e)\, e^j$, contract with a smearing field $f_{ij}$, and integrate over $\Sigma$ to find\relax
\footnote{We include the initial sign factor in the definition of $\mbb{B}[f]$ because the integrand in the second factor is a 3-form.  A 3-form can be integrated over an \textit{oriented} manifold $\Sigma$, and the sign of the integral switches if that  orientation is reversed.  Thus, the initial sign factor makes $\mbb{B}[f]$ independent of the orientation of $\Sigma$.  The proper-volume integral $V[f]$ of $\tr f$ on the right is manifestly orientation-independent, so this sign is needed to make a meaningful  comparison.}
\begin{equation}\label{SC:cBdef}
                \mbb{B}[f] := \sgn(\det e) \int_\Sigma \mbb{F}^i\! \wedge e^j\, f_{ij}
                                = \mbb{b} \int_\Sigma f_i{}^i\, \abs{\det e} =: \mbb{b}\, V[f].
\end{equation}
Demanding this condition for all $f_{ij}$ is equivalent to (\ref{SC:sComD}).  Meanwhile, the non-complexified version $B[f]$ of the left side is closely related to the Euclidean (self-dual) Hamiltonian constraint $C_{\mathrm{E}}[N]$.  Specifically, we  have $B[f] = 3 C_{\mathrm{E}}[N]$ for $f_{ij} = N\, \delta_{ij}$.  One can therefore mimic Thiemann's technique \cite{thiemann1996} to regularize and quantize the Euclidean Hamiltonian constraint to quantize $B[f]$, and then complexify the resulting operator  as described above.  Finally, since $\mbb{b}$ can take \textit{any} value, we cross-multiply two instances of (\ref{SC:cBdef}) to write the symmetry conditions in the implicit form
\begin{equation}\label{SC:cSdef}
                \mbb{S}[f, g] := \mbb{B}[f]\, V[g] - V[f]\, \mbb{B}[g] = 0.
\end{equation}
Choosing $g_{kl} = \delta_{kl}$ to be the identity gives $V[1] = 3 V > 0$, so we can always solve this equation for
\begin{equation}\label{SC:cSequiv}
                \mbb{B}[f] = \frac{\mbb{B}[1]}{3V}\, V[f] + \frac{\mbb{S}[f, 1]}{3V}.
\end{equation}
Setting $\mbb{S}[f, 1] = 0$ for all $f_{ij}$ therefore implies (\ref{SC:cBdef}) with $\mbb{b} = \mbb{B}[1] / 3 V$.   Thus, the implicit form (\ref{SC:cSdef}) of the symmetry conditions implies all of the prior versions of those conditions discussed above.   In other words, the symmetric sector of the classical theory consists precisely of the intersection of the submanifolds $\mbb{S}[f, g] = 0$ in phase space over all pairs of smearing functions $(f_{ij}, g_{kl})$.  Moreover, up to the usual ordering ambiguities,  we have sketched how it is possible to construct analogous operators $\hat{\mbb{S}}[f, g]$ in the quantum theory.

\subsection{Poisson algebra of the symmetry conditions}
\label{SC.palg}
 
As discussed in the preamble to this section, although demanding $\mbb{S}[f, g] = 0$ for all pairs of smearing functions $(f_{ij}, g_{kl})$ selects the homogeneous and isotropic sector of the classical theory, this does not imply that the analogous conditions  $\hat{\mbb{S}}[f, g]\, \ket{\psi} = 0$ can be imposed simultaneously for all pairs of smearing fields to select symmetric quantum states.  A necessary, though generally not sufficient, condition for this to be possible is that the Poisson algebra of the $\mbb{S}[f,  g]$ closes in the classical theory.  We show in this subsection that it does.
 
We calculate the Poisson algebra of the symmetry conditions in several steps.  First, define non-complexified analogues of the quantities appearing in (\ref{SC:cBdef}), setting
\begin{equation}\label{SC:BVdefs}
                B[f] := \sgn(\det e) \int_\Sigma f_{ij}\, F^i \wedge e^j
                \qquad\text{and}\qquad
                V[f] := \int_\Sigma \tr f\, \abs{\det e}.
\end{equation}
The Poisson algebra of the phase-space functionals (\ref{SC:cSdef}) derives first of all from
\begin{equation}\label{SC:rBBbra}
                \pbrack[\big]{B[f]}{B[g]} = \newton\bip\, B[f\, \rKop g - g\, \rKop f],
\end{equation}
where $(f\, \rKop g)_{ik}$ denotes the matrix product of $f_{ij}$ and
\begin{equation}\label{SC:rKdef}
                (\rKop g)^j{}_k := \frac{e^j \wedge D(g_{kl}\, e^l)}{\svol(e)} - \delta^j_k\, \frac{e^i \wedge D(g_{il}\, e^l)}{2 \svol(e)}.
\end{equation}
The first-order (but not Leibniz) differential operator $\rKop$ acting on smearing fields involves the \textit{signed} volume element $\svol(e) := \epsilon_{ijk}\, e^i\, e^j\, e^k$ induced on $\Sigma$ by the triad $e^a_i$.  (The ratio of 3-forms is well-defined  because the space of 3-forms is one-dimensional.)  The other basic Poisson bracket we will need is
\begin{align}\label{SC:rBVbra}
                \pbrack[\big]{B[f]}{V[g]}
                                &= \frac{\newton\bip}{2} \sgn(\det e) \int \tr g\, e^i \wedge D (f_{ij}\, e^j)
                                                \notag\\
                                &= - \newton\bip\, V \bigl[ (\tr g)\, \rKop f \bigr].
\end{align}
These preliminary results yield the Poisson bracket
\begin{equation}\label{SC:rSSbra}
                \pbrack[\big]{S[f, g]}{S[h, k]}
                                = \newton\bip\, \Bigl[ \Bigl( V[g]\, S \bigl[ f\, \rKop h - h\, \rKop f, k \bigr] - S[g, k]\, V \bigl[ (\tr f)\, \rKop h \bigr] \Bigr) - (f \leftrightarrow g) \Bigr] - [h \leftrightarrow k]
\end{equation}
of the non-complexified analogues of the symmetry conditions (\ref{SC:cSdef}).  Here we have used the identity
\begin{equation}\label{SC:rKibp}
                V \bigl[ f\, \rKop g - g\, \rKop f \bigr] = V \bigl[ (\tr f)\, \rKop g - (\tr g)\, \rKop f \bigr],
\end{equation}
which arises, at least for smearing functions of compact support, because the first terms from (\ref{SC:rKdef}) on the left side combine to give an exact exterior derivative.
 
Note that each of the eight terms on the right side of (\ref{SC:rSSbra}) is proportional to one of the (non-complexified) symmetry conditions.  The Poisson algebra of the \textit{real} symmetry conditions therefore closes.  But we have argued above that  the complexification process commutes both with products and with Poisson brackets.  It follows immediately that the complexified symmetry conditions satisfy
\begin{equation}\label{SC:cSSbra}
                \pbrack[\big]{\mbb{S}[f, g]}{\mbb{S}[h, k]}
                                = \newton\bip\, \Bigl[ \Bigl( V[g]\, \mbb{S} \bigl[ f\, \cKop h - h\, \cKop f, k \bigr] - \mbb{S}[g, k]\, V \bigl[ (\tr f)\, \cKop h \bigr] \Bigr) - (f \leftrightarrow g) \Bigr] - [h \leftrightarrow k],
\end{equation}
where we have introduced the complexified analogue
\begin{equation}\label{SC:cKdef}
                (\cKop g)^j{}_k
                                := \frac{e_j \wedge \mbb{D}(g_{kl}\, e^l)}{\svol(e)} - \delta^j_k\, \frac{e^i \wedge \mbb{D}(g_{il}\, e^l)}{2 \svol(e)}
                                = (\rKop g)^j{}_k - \ii\alpha\, g^j{}_k
\end{equation}
of the differential operator from (\ref{SC:rKdef}).  Thus, the classical Poisson algebra of the complexified symmetry conditions closes.  Barring anomalies arising from quantization, the conditions $\hat{\mbb{S}}[f, g] \ket{\psi} = 0$ for all pairs of  smearing fields $(f_{ij}, g_{kl})$ are therefore consistent with one another, and may be applied simultaneously to select symmetric states $\ket{\psi}$ at the quantum level.
 
Let us highlight a particularly important point here.  The right side of (\ref{SC:cSSbra}) would still vanish on the space of classical symmetric states if it included terms proportional to complex conjugates $\bar{\mbb{S}}[f, g]$ of the symmetry conditions.   The Poisson algebra would then close only if we expanded it to include the complex conjugate functions $\bar{\mbb{S}}[f, g]$.  But the \emph{quantum} constraints would then be inconsistent with one another since the Poisson bracket of $\bar{\mbb{S}}[f, g]$  and $\mbb{S}[h, k]$ does not vanish, even if the symmetry conditions hold.

Even without explicitly calculating Poisson brackets, we know this latter fact must be true because the Poisson algebra spanned by $\mbb{S}[f, g]$ and $\bar{\mbb{S}}[f, g]$ includes 
$\real\mbb{S}[f, g]$ and $\imag\mbb{S}[f, g]$, which form a set of real-valued functionals that together constrain to vanish the 
non-symmetric components of conjugate pairs of configuration and momentum degrees of freedom in the classical phase space.
As a consequence, $\real \mbb{S}[f, g]$ and $\imag \mbb{S}[f, g]$ necessarily form a Poisson algebra that is second class in the sense defined by Dirac for constraints \cite{dirac1964}, and must have Poisson brackets that do not vanish in the symmetric  sector.  Hence, there must also be Poisson brackets among the  $\mbb{S}[f, g]$ and $\bar{\mbb{S}}[f, g]$ that do not vanish on the 
symmetric sector.

For completeness, let us verify this by precisely because homogeneity is imposed classically extending the Poisson algebra of the $\mbb{S}[f, g]$ found above to include their complex conjugates $\bar{\mbb{S}}[f, g]$.  To do this, first observe that (\ref{SC:comF})  gives
\begin{align}\label{SC:rYdef}
                \mbb{B}[f] = B[f] + \ii\alpha\, Y[f] - \alpha^2\, V[f]
                \qquad\text{with}\qquad
                Y[f] := \frac{2}{\newton\bip}\, \pbrack[\big]{B[f]}{V}
                                &= \sgn(\det e) \int_\Sigma e^i \wedge D(f_{ij}\, e^j)
                                                \notag\\
                                &= -2\, V[\rKop f].
\end{align}
Extending the Poisson algebra from (\ref{SC:rBBbra}) and (\ref{SC:rBVbra}) to include the phase-space functionals $Y[f]$ gives
\begin{align}\label{SC:YBbra}
                \pbrack[\big]{Y[f]}{B[g]} &= \newton\bip\, \bigl( Y[f\, \rKop g] + B[g f] + V[(\rKop f - 1 \tr \rKop f)\, \rKop g] \bigr),
                                \\[1ex]\label{SC:YVbra}
                \pbrack[\big]{Y[f]}{V[g]} &= \newton\bip\, V[(\tr f)\, g],
\intertext{and}\label{SC:YYbra}
                \pbrack[\big]{Y[f]}{Y[g]} &= - \newton\bip\, Y[fg - gf].
\end{align}
Since the term at order $\alpha^2$ in $\mbb{B}[f]$ is proportional to $V[f]$, we have simply
\begin{equation}\label{SC:rTdef}
                \mbb{S}[f, g] = S[f, g] + \ii\alpha\, T[f, g]
                \qquad\text{with}\qquad
                T[f, g] := Y[f]\, V[g] - V[f]\, Y[g]
\end{equation}
Using the preliminary Poisson brackets computed above, the complete Poisson algebra of the classical symmetry conditions can be expressed in terms of (\ref{SC:rSSbra}), together with
\begin{align}\label{SC:rSTbra}
                \pbrack[\big]{S[f, g]}{T[h, k]}
                                &= - \newton\bip\, \smash[b]{\Bigl[ \Bigl(}
                                                V[g]\, T[h\, \rKop f, k]
                                                + V[k]\, S[fh, g]
                                                + V[g]\, V[k]\, V \bigl[ (\rKop f - 1 \tr \rKop f)\, \rKop h \bigr]
                                                \\&\hspace{2em}\notag
                                                + B[g]\, V[k]\, V \bigl[ (f - 1 \tr f)\, h \bigr]
                                                + V[g]\, Y[k]\, V \bigl[ (f - 1 \tr f)\, \rKop h \bigr]
                                                \smash[t]{\Bigr) - (f \leftrightarrow g) \Bigr]} - [h \leftrightarrow k],
\intertext{and}\label{SC:rTTbra}
                \pbrack[\big]{T[f, g]}{T[h, k]}
                                &= - \newton\bip\, \Bigl[ \Bigl(
                                                                V[g]\, T[fh - hf, k]
                                                                - T[g, k]\, V \bigl[ (\tr f)\, h \bigr]
                                                                \Bigr) - (f \leftrightarrow g) \Bigr] - [h \leftrightarrow k].
\end{align}
The last three terms in braces in (\ref{SC:rSTbra}) are symmetric under the interchange $f \leftrightarrow h$ of smearing functions, where the symmetry of the last term follows from (\ref{SC:rKibp}).  This observation, together with the identity
\begin{equation}
                V[f]\, S[g, h] + V[g]\, S[h, f] + V[h]\, S[f, g] = 0,
\end{equation}
facilitates a direct calculation confirming (\ref{SC:cSSbra}).  These results also give the bracket
\begin{align}\label{SC:cSS*bra}
                \pbrack[\big]{\mbb{S}[f, g]}{\bar{\mbb{S}}[h, k]}
                                &= \newton\bip\, \smash[b]{\Bigl[ \Bigl(}
                                                V[k]\, \mbb{S} \bigl[ f\, \bar{\cKop} h, g \bigr]
                                                - V[g]\, \bar{\mbb{S}} \bigl[ h\, \cKop f, k \bigr]
                                                                \notag\\&\hspace{4em}
                                                + \mbb{B}[g]\, V[k]\, V \bigl[ (f - 1 \tr f)\, \bar{\cKop} h \bigr]
                                                - V[g]\, \bar{\mbb{B}}[k]\, V \bigl[ (f - 1 \tr f)\, \cKop h \bigr]
                                                                \notag\\&\hspace{4em}
                                                + 2 \ii\alpha\,  V[g]\, V[k]\, \real V \bigl[ (\bar{\cKop} f - 1 \tr \bar{\cKop} f)\, \cKop h \bigr]
                                                \smash[t]{\Bigr) - (f \leftrightarrow g) \Bigr]} - [h \leftrightarrow k]
\end{align}
between the complex symmetry conditions and their complex conjugates.
 
Now consider the bracket (\ref{SC:cSS*bra}) at a background point lying in the maximally symmetric submanifold of the classical phase space.  At such a point we have
\begin{equation}
\label{eqn:hired}
                \mbb{B}[f] \approx \bigl( \rho + (\bip H + \ii \alpha)^2 \bigr)\, V[f]       
                \qquad\text{and}\qquad
                (\mbb{K} f)_{ij} \approx ({}^\Gamma \rKop f)_{ij} - (\bip H + \ii\alpha)\, f_{ij},
\end{equation}
where ${}^\Gamma \rKop$ is the operator (\ref{SC:rKdef}) constructed from the spin connection.  It follows that (\ref{SC:cSS*bra}) reduces to 
\begin{align}
                \pbrack[\big]{\mbb{S}[f, g]}{\bar{\mbb{S}}[h, k]}
                                &\approx 2 \ii \alpha \newton\bip\, \smash[b]{\Bigl[ \Bigl(}
                                                V[g]\, V[k]\, V \bigl[ \bigl( {}^\Gamma \rKop f - 1 \tr {}^\Gamma \rKop f \bigr)\, {}^\Gamma \rKop h \bigr]
                                                                \notag\\&\hspace{8em}
                                                + \rho\, V[g]\, V[k]\, V \bigl[ (f - 1 \tr f)\, h \bigr]
                                                \smash[t]{\Bigr) - (f \leftrightarrow g) \Bigr]} - [h \leftrightarrow k].
\end{align}
Now, choosing $f_{ij} = h_{ij} = \phi\, \delta_{ij}$ to be pure trace with $\phi$ variable and $g_{ij} = k_{ij} = n\, \delta_{ij}$ to be pure trace with $n$ constant, one finds that
\begin{equation}
                \pbrack[\big]{\mbb{S}[\phi 1, n 1]}{\bar{\mbb{S}}[\phi 1, n 1]} = - 36 \ii\alpha\kappa\bip n^2 V^2 \int_\Sigma \Bigl( \norm[\big]{\ed\phi}^2 +  3 \rho\, \bigl( \phi - \expect{\phi} \bigr)^2 \Bigr)\, \abs{\det e}.
\end{equation}
Clearly one can choose $\phi$ such that the right side is non-zero.  This counterexample shows explicitly that the extended Poisson algebra of the complexified symmetry conditions $\mbb{S}[f, g]$ and their complex conjugates $\bar{\mbb{S}}[h, k]$ does  not close.

\subsection{Inclusion of constraints}
\label{SC.dynam}
 
The kinematical (\textit{i.e.}, Gauss and diffeomorphism) constraints are solved at the classical level in loop quantum cosmology, prior to quantization, by fixing an action of the symmetry group as described above.  Accordingly, any comparison between  loop quantum cosmology and the symmetric sector of loop quantum gravity ought to occur after those kinematical constraints have been solved in the full theory.  Since the symmetry conditions (\ref{SC:cSdef}) are gauge-invariant and diffeomorphism-covariant,  however, it follows immediately that
\begin{equation}
                \pbrack[\big]{C[\Lambda]}{\mbb{S}[f, g]} = 0
                \qquad\text{and}\qquad
                \pbrack[\big]{C[\vec N]}{\mbb{S}[f, g]} = \mbb{S}[\Lie_{\vec N} f, g] + \mbb{S}[f, \Lie_{\vec N} g],
\end{equation}
where $C[\Lambda]$ and $C[\vec N]$ denote the smeared Gauss and diffeomorphism constraints, respectively.  There is therefore no (classical) obstruction to imposing the (quantum) 
symmetry conditions within the space of states that solve the Gauss and diffeomorphism constraint.
 
Furthermore, the Hamiltonian constraint in the cosmological model is a single condition $C = 0$ in the reduced phase space, whereas in the full theory it consists of an infinite-dimensional family $C[N] = 0$ of distinct conditions.  However, the smeared  constraint functions $C[N]$ are redundant when restricted to the homogeneous subspace of the phase space of general relativity.  It is straightforward to show that
\begin{equation}
                C[N] \equiv \frac{C[1]\, V[N]}{V}
\end{equation}
on the submanifold of symmetric classical states, where $V[N] := \int_\Sigma N\, \ed V$ denotes the proper-volume integral of the lapse over all space.  Accordingly, the Hamiltonian constraint we will seek to impose in the symmetric sector of loop quantum  gravity, and to compare with the Hamiltonian constraint of loop quantum cosmology, will correspond to $C[1]$.
This choice of lapse is also convenient as it yields a Hamiltonian constraint which is diffeomorphism invariant
and therefore is expected to give rise to a quantum constraint which preserves the space of solutions to both the diffeomorphism constaint and the Gauss constraint.
 
The Poisson algebra of the symmetry conditions $\mbb{S}[f, g]$ and the \textit{Euclidean} Hamiltonian constraint $C_{\text{E}} = \frac{1}{3}\, B[1]$ for constant lapse is straightforward to compute using the results above.  One finds that
\begin{align}
\label{SC:SHbra}
                \pbrack[\big]{\mbb{S}[f, g]}{C_{\mathrm{E}}}
%                             &= \newton\bip\, \smash[b]{\Bigl(}
%                                             \mbb{S}[f\, \rKop n, g]
%                                             + \bigl( B[g] + \ii\alpha\, Y[g] \bigr)\, V[(n - 1 \tr n)\, \rKop f]
%                                             + \ii\alpha\, B[n f]\, V[g]
%                                                             \notag\\&\hspace{4em}
%                                             - B[n\, \rKop f]\, V[g]
%                                             + \ii\alpha\, V[(n - 1 \tr n)\, \rKop^2 f]\, V[g]
%                                             \smash[t]{\Bigr)} - (f \leftrightarrow g)
%                                                             \notag\\
%                             &= \newton\bip\, \smash[b]{\Bigl(}
%                                             \mbb{S}[f\, \rKop n, g]
%                                             - \bar{\mbb{S}}[n\, \rKop f, g]
%                                             + \tfrac{3}{2}\, n\, Y[f]\, B[g]
%                                             + \ii\alpha n\, B[f]\, V[g]
%                                             \smash[t]{\Bigr)} - (f \leftrightarrow g)
%                                                             \notag\\
                                &= \frac{\newton\bip}{3}\, \biggl(
                                                \mbb{S}[f\, \rKop 1, g]
                                                - \bar{\mbb{S}}[\rKop f, g]
                                                + \frac{\mbb{S}[f, 1]\, \bar{\mbb{S}}[g, 1]}{12 \ii \alpha V^2}
                                                \biggr) - (f \leftrightarrow g)
                                                \notag\\&\hspace{4em}
                                                + \newton\bip\, C_{\text{E}}\, \frac{\mbb{S}[f, g] - \bar{\mbb{S}}[f, g]}{4 \ii\alpha\, V}
                                                - \newton\bip\, \bigl( \ii\alpha\, Y[1]\, + 6 \alpha^2\, V \bigr)\, \frac{\mbb{S}[f, g] + \bar{\mbb{S}}[f, g]}{12 \ii\alpha V}.
\end{align}
As one would expect on physical grounds, the right side vanishes on the submanifold of classical symmetric states,
since the symmetry is preserved under classical time evolution.  
However, the Poisson algebra of symmetry conditions and the Hamiltonian constraint only closes if we include the complex conjugate symmetry conditions $\bar{\mbb{S}}[f, g]$.  That is, we cannot expect the Hamiltonian constraint \textit{operator} 
to preserve the \textit{quantum} symmetric sector.  
Consequently, comparison of the full theory Hamiltonian constraint with that in LQC will need to involve, \textit{e.g.}, 
the matrix elements of the former.  Though we have not checked it explicitly, we see no reason to expect the situation would be any better for the Lorentzian Hamiltonian constraint.

\section{Quantization: The quantum symmetric sector}

\subsection{Quantization of the symmetry conditions}

Let us now proceed to show how the diffeomorphism-invariant homogeneous isotropic condition
(\ref{SC:cBdef}) can be quantized. One proceeds in two steps: First, define
an operator corresponding to each quantity $\mathbb{B}[\matdec{f}]$ and then use this to define
$\hat{\mathbb{S}}[f,g]$ and impose the quantum analogue of (\ref{SC:cBdef}).
Following (\ref{SC:rYdef}), we set 
\begin{equation}\label{qBcomplex}
\hat{\mbb{B}}[f] := \hat B[f] + \ii\alpha\, \hat Y[f] - \alpha^2\, \hat V[f] 
\qquad\text{with}\qquad 
\hat Y[f] := \frac{2}{\ii\hbar\kappa\gamma}\, \comm[\big]{\hat B[f]}{\hat V}, 
\end{equation}
where we now need to specify the operator $\hat B[f]$.

The form of $B[f]$ in (\ref{SC:BVdefs}) is almost identical to that of the Euclidean self-dual Hamiltonian constraint,
and the exact same methods can be used to quantize it \cite{thiemann1996}.
No new quantization procedures need be invented.
Specifically, one can use the standard so-called `Thiemann trick' to write
$B[f]$ in terms of only curvature, the connection, and the volume of the universe: 
\begin{align*}
B[f] = \int f^i{}_j F^j \wedge e_i
= \frac{2}{\newton\gamma} \int f^i{}_j F^j \wedge \{A_i, V\} .
\end{align*}
%
% I think we ended up not doing the following; I'm not 100% sure.
% 
%\begin{align*}
%B[f] = \sgn(\det e) \int f^i{}_j F^j \wedge e_i
%= 2 \sgn(\det e) \int f^i{}_j F^j \wedge \{A_i, V\} .
%\end{align*}
% In \cite{thiemann2007}, the $\sgn(\det e)$ factor is also present and is absorbed into the lapse, 
% making it a pseudo scalar.
%
A regulated version $B[f]_\epsilon$ of this expression can be constructed,
with $F$ and $A$ represented by closed and open holonomies
exactly as is done for the Hamiltonian constraint \cite{thiemann1996, al2004, thiemann2007}.
Everything in the resulting expression for $B[f]_\epsilon$ has a direct quantum analogue,
with $V$ quantized in the standard way \cite{rs1993, al1997}, and
the Poisson bracket quantized as a commutator, leading to an operator $\hat{B}[f]_\epsilon$ on $\Cyl$.
%
% Below gives more detail from my presentation of the Hamiltonian constraint in my piecewise linear LQG paper.
% I decided to not include this.
%
%For each lapse $N$, each $\epsilon \in [0,\epsilon_0]$
%%
%% alrev also says just this, without defining $\epsilon_0$, and it is
%% clear what is meant, so I will do the same (and I am intending to keep this to
%% minimum).
%%
%and each graph $\gamma$,
%one defines a regulated operator $\hat{C}[N]_{\gamma, \epsilon}$ on $\Hort{\gamma}$
%(see \cite{thiemann1996, al2004}).
% Piecing these together for all $\gamma$ gives, for each $\epsilon$, an operator
%$\hat{C}(N)_\epsilon$ on the kinematical Hilbert space $\Hil$.
%
The dual $\hat{B}[f]_\epsilon^*$ then acts on $\Cyl^{\star}$:
$\hat{B}[f]_\epsilon^* \big[ (\Psi| \big] = (\Psi | \hat{B}[f]_\epsilon$.
For any $\Psi$ in
$\Cyl_\diff^{\star} \subset \Cyl^{\star}$, the limit
$\lim_{\epsilon \rightarrow 0} (\Psi| \hat{B}[f]_\epsilon$
becomes trivial exactly in the same manner as for the Hamiltonian constraint
\cite{thiemann1996, rs1993}, allowing us to define
\begin{equation}
(\Psi| \hat{B}[f] := \lim_{\epsilon \rightarrow 0} (\Psi|\hat{B}[f]_\epsilon ,
\end{equation}
so that $\hat{B}[f]$ is well-defined on $\Cyl_\diff^{\star}$.
With $\hat{B}[f]$ defined, equation (\ref{qBcomplex}) gives us the operator $\hat{\mathbb{B}}[f]$.
The resulting operator $\hat{\mathbb{B}}[f]$ is \textit{diffeomorphism covariant}, as must be the
case from the background independence of its construction.
%
% In the sense that, for any diffeomorphism $\varphi$,
% $\widehat{\mathbb{B}[\varphi^*f]} = U_\varphi \widehat{\mathbb{B}[f]} U_{\varphi}^{-1}$
% (I'm not sure of which action of $\varphi$ on the RHS should be an inverse.)
%
However, again similar to the Hamiltonian constraint operator,
for general smearing function $f^i{}_j$, $\hat{\mathbb{B}}[f]$ will map $\Cyl_\diff^{\star}$ out of
itself due to $\hat{\mathbb{B}}[f]$ not being diffeomorphism \textit{invariant}.

With $\hat{\mathbb{B}}[f]$ defined, it remains only to define the  quantization of the smeared volume $V[f]$.
 But this is easy, because, in fact,
the usual volume operator in loop quantum gravity takes the form $\hat{V} = \sum_x \sqrt{|\widehat{h_x}|}$
\cite{al1997, gt2005}, so that one has an operator-valued distribution corresponding to the volume element
$\widehat{\sqrt{h(x)}}:= \sum_v \delta^3(v,x) \sqrt{|\widehat{h_v}|}$ \cite{al1997}, which leads to
\begin{align*}
\hat{V}[f]:= \sum_x f(x)^i{}_i \sqrt{|\widehat{h_x}|}.
\end{align*}
The quantization of $\mathbb{S}[f,g]$ (\ref{SC:cSdef}) then gives
\begin{align*}
\hat{\mathbb{S}}[f,g] &:= \hat{\mathbb{B}}[g]  \, \hat{V}[f]
- \hat{\mathbb{B}}[f] \, \hat{V}[g].
\end{align*}
The symmetric sector of diffeomorphism invariant LQG, which we denote $\Vsymm$,
is then defined to be the set of all $\Psi \in \Cyldiff$
satisfying a quantization of (\ref{SC:cSdef}):
\begin{align}
\label{eqn:qsymmsect}
(\Psi |  \hat{\mathbb{S}}[f,g] = 0
\end{align}
for all $f^i{}_j$ and $g^i{}_j$.
Note that, because our construction of $\hat{\mathbb{B}}[f]$ has well-defined action only on diffeomorphism invariant 
states, $\hat{\mathbb{S}}[f,g]$ too acts only on diffeomorphism invariant states. 
Consequently, we must look for solutions to the quantum symmetry condition in the space of diffeomorphism invariant states.
These are not normalizable, 
and so must be represented in $\Cyl^\star$, whence (\ref{eqn:qsymmsect}) is the desired mathematically precise version of 
the quantum symmetry constraint.
%$\hat{\mathbb{S}}[f,g] |\psi \rangle = 0$ .

\subsubsection*{Remark:}

As noted, $\mathbb{B}[f]$ is the complexification of $B[f]$ with complexifier $V/v_o$. As a consequence, 
as an alternative to (\ref{qBcomplex}), $\mathbb{B}[f]$ can be quantized as
\begin{align}
\label{qBcomplexify}
\hat{\mathbb{B}}'[f] := e^{-\hat{V}/v_o} \hat{B}[f] e^{\hat{V}/v_o} .
\end{align}
This will yield the same operator $\hat{\mathbb{B}}[f]$ in (\ref{qBcomplex}) if the 
Poisson bracket relation $\{\{ B[f], V\}, V\} = \frac{\newton^2 \gamma^2}{2} V[f]$ does not develop an anomaly during quantization: 
\begin{align}
\label{dcommrel}
\left[\left[\hat{B}[f], \hat{V}\right], \hat{V} \right] = \frac{-\hbar^2 \newton^2 \gamma^2}{2} \hat{V}[f].
\end{align}
Yet a third alternative is to directly quantize (\ref{SC:cBdef}), 
using the quantization (\ref{SC:chop}) of holonomies of $\mathbb{A}^i_a$ and again using methods similar to those
used in the quantization of the Thiemann constraint.
This result will be the same as (\ref{qBcomplexify}): because $\hat{V}$ commutes with all the factors besides the holonomies,
all $e^{\hat{V}}$ and $e^{-\hat{V}}$ factors in the middle of the expression cancel,
and one is left with equation (\ref{qBcomplexify}).
% which is simpler.

The expression (\ref{qBcomplex}), however, is by far the simplest to actually use in calculations, and we therefore advocate it in the case where these two quantizations of $\mathbb{B}[f]$ are inequivalent --- \textit{i.e.}, in the case where (\ref{dcommrel})  does not hold.

\subsection{Average spatial curvature operator}
\label{subsect:curv}

We will be interested in embedding a specific LQC minisuperspace model into
the symmetric sector $\Vsymm$.  However, specific LQC models always
restrict to one of the three cases of zero, positive, or negative spatial scalar curvature,
referred to as $k=0$, $k=1$, and $k=-1$, respectively.
Because of this, it is important to have a tool to be able to further distinguish, within
$\Vsymm$, these different signs of the spatial curvature, to ensure that one
is embedding into the correct $k$ sector in the full theory.  To this end,
we construct an operator encoding the spatial
scalar curvature as follows.
The spatial scalar curvature $R$ is none other than 6 times the constant $\rho$
appearing in equation (\ref{SC:sAshD}) \cite{carroll2013}.
From equation
(\ref{eqn:hired}) one then obtains, in the homogeneous isotropic sector,
\begin{align*}
R &\approx
\frac{\Re \mathbb{B}[\ident]}{3V} + \alpha^2
- \left(\frac{\Im \mathbb{B}[\ident]}{6 \alpha V}\right)^2
\end{align*}
where $V$ is the total spatial volume.
% of the universe.
The right hand side of the above equation, when no longer restricted to the homogeneous isotropic sector,
provides a notion of `averaged' spatial scalar curvature $R_{\text{ave}}$.
Multiplying this expression by $V^2$ removes all the volumes from the denominator,
\begin{align}
\label{RaveVc}
R_{\text{ave}} V^2 := \frac{\Re \mathbb{B}[\ident] V}{3} + \alpha^2 V^2
- \left(\frac{\Im \mathbb{B}[\ident]}{6\alpha}\right)^2.
\end{align}
Substituting in the expressions for $\Re \mathbb{B}[1]$ and $\Im \mathbb{B}[1]$ from 
(\ref{SC:rYdef}),
\begin{align}
R_{\mathrm{ave}} V^2 = \frac{B[1] V}{3} - \left(\frac{Y[1]}{6}\right)^2 .
\end{align}
This expression is readily quantized, yielding an operator defined on all of $\CylDiff$:
\begin{align}
\label{RaveVq}
\widehat{R_{\text{ave}}V^2} =
\frac{1}{6}\left(\hat{\mathrm{B}}[\ident] \hat{V}+  \hat{V} \hat{\mathrm{B}}[\ident]\right)
- \left(\frac{\hat{Y}[\ident]}{6}\right)^2
\end{align}
where the first term is symmetrically ordered in order to ensure that the resulting operator is self-adjoint.

\section{Embedding strategy}
% From this point on, we are presenting a mix of results and ``strategy'', whereas before this point
% everything is just results. Not only is the presentation natural this way, but  keeping these parts separated like this
% strengthens the presentation of the of the definitiveness of the first part, which is impressive in its own right,
% *especially in light of the second section on strategy, which is equally impressive.*

\subsection{Choosing an embedding}

\label{subsect:iotadef}

Simply defining a diffeomorphism-invariant, homogeneous isotropic sector is not enough: We wish to additionally relate this sector
to loop quantum cosmology.  This is accomplished by additionally defining an \textit{embedding} of LQC states
into the diffeomorphism-invariant, homogeneous isotropic sector.

We wish to find an embedding $\iota$ from LQC states $\Hil_S$ into the symmetric sector $\Vsymm$.
In other words, we wish to find an embedding $\iota: \Hil_S \rightarrow \Cyldiff$
satisfying
\begin{align}
\label{iotasymm}
\hat{\mathbb{S}}[f,g] \circ \iota = 0
\end{align}
for all $f,g$.
However, this will clearly not be enough to determine $\iota$ uniquely.  Indeed, this equation demands only that the image of $\iota$ is a subspace of the symmetric sector in LQG.  It does not specify which subspace (recall that, at a minimum, the symmetric  sector in LQG contains states corresponding to all three signs $k = 0, \pm 1$ of the spatial curvature), nor the exact correspondence of LQC states with LQG states in that subspace.  From a simple counting
argument (seen most clearly in the toy example in  \appref{toyapp}), one expects
to be able to impose two more conditions on $\iota$ --- corresponding to the two dimensions of the
symmetry reduced phase space.
As clarified in the toy example in \ref{toyapp},
these conditions can take one of two forms:
in terms of \textit{intertwining} or \textit{equality of matrix elements}
of certain operators.
If $\hat{O}_1$, $\hat{O}_2$ are two operators of particular interest in the full theory,
and $\hat{O}_1^S$, $\hat{O}_2^S$ the corresponding LQC operators, 
these two possibilities take the following forms:

\noindent (1.) Intertwining:
\begin{align}
\label{intertwine}
\iota \circ \hat{O}_i^S = \hat{O}_i \circ \iota, \quad i=1,2.
\end{align}
%
% Note: Sometimes I am confused about a potential subtlety in the above equation, namely, whether, because
% $\iota$ is anti-linear and maps kets to bras, one side of the above equation should have an adjoint on it, or
% a dual.
%
% This question is made much simpler because we have defined the linear structure on $\Cyl^\star$ to
% be opposite that which it usually is, so that $\iota$ is linear.  With this convention (see preliminaries),
% any adjoint or dual is actually part of the definition of $\widehat{O}_i$ in the full theory when acting
% on $\Cyl^\star$, so that the above equation is exactly correct.
%
or (2.) Equality of matrix elements:
\begin{align}
\label{eqn:matel}
\frac{\langle \iota \psi, \hat{O}_i \iota \phi \rangle}{\sqrt{\langle \iota \psi, \iota \psi \rangle
\langle \iota \phi, \iota \phi \rangle}} 
= \frac{\langle \psi, \hat{O}_i^S \phi \rangle}{\sqrt{\langle \psi, \psi \rangle \langle \phi, \phi \rangle}}
\quad \text{for all } \psi, \phi \in \Hil_S .
\end{align}
The division by the norms here is needed because $\iota$ will in general not be norm-preserving, so that $\phi$, $\psi$,
$\iota \phi$, $\iota \psi$ will in general not be simultaneously normalizable.

For the operators $\hat{O}_i$, $\hat{O}_i^S$ chosen, the \textit{existence} of an $\iota$
satisfying (\ref{intertwine}) or (\ref{eqn:matel})
enforces first a non-trivial consistency condition on the
quantization of the $\hat{O}_i$ in the reduced and full theories.
Once $\iota$ is fixed through the two conditions (\ref{intertwine}),
one can then use it to compare \textit{other} operators in the reduced and full theories.
If another operator $\hat{O}$,
preserves $\Vsymm$, there will exist an operator $\hat{O}_S$
on $\Hil_S$ satisfying
\begin{align}
\hat{O} \circ \iota = \iota \circ \hat{O}_S,
\end{align}
This equation determines $\hat{O}_S$ uniquely so that $\hat{O}_S$ is \textit{induced} on $\Hil_S$
via this condition.  If $\hat{O}$ does not preserve $\Vsymm$, then an operator $\hat{O}_S$
can be induced through matrix elements:
\begin{align}
\frac{\langle \iota \psi, \hat{O} \iota \phi \rangle}{\sqrt{\langle \iota \psi, \iota \psi \rangle \langle \iota \phi, \iota \phi \rangle}} 
= \frac{\langle \psi, \hat{O}_S \phi \rangle}{\sqrt{\langle \psi, \psi \rangle \langle \phi, \phi \rangle}}
\quad \text{for all } \psi, \phi \in \Hil_S .
\end{align}
In either case, the induced operator $\hat{O}_S$
may or may not match the quantization of the same quantity used in LQC thus far.
If they do not match,
one can then discuss possible modifications to the quantization of
$\hat{O}$ in the full and reduced theory to achieve agreement.

The equations (\ref{iotasymm}, \ref{intertwine}) or (\ref{iotasymm}, \ref{eqn:matel})
can be solved by taking components in the momentum
$\{|p\rangle\}$ and spin-network
$\{|\gamma, \vec{j}, \vec{T} \rangle \}$
bases of $\Hil_S$ and $\Hil$, respectively. Equations (\ref{iotasymm}, \ref{intertwine})
or (\ref{iotasymm}, \ref{eqn:matel})
then reduce to a set of difference equations for the matrix
elements $( \iota p | \gamma, \vec{j}, \vec{T} \rangle$
of $\iota$.  Because $\iota$ maps into diffeomorphism invariant states, these matrix elements depend on
$(\gamma,\vec{j}, \vec{T})$ only via the diffeomorphism equivalence class $[(\gamma,\vec{j}, \vec{T})]$.
The lessons of \cite{engle2006, engle2006a}, the application of this strategy to embedding into Bianchi I \cite{behm2017},
as well as  the toy example discussed in  \appref{toyapp}
each give reason to believe that the above equations will uniquely determine the matrix elements  $( \iota p | [\gamma, \vec{j}, \vec{T}] \rangle$.
%
% and not over-determine.
%

Alternatively, one or both of these conditions might be replaced by a condition with an equivalent amount
of information.  We will discuss one possible choice of operator to intertwine in subsection \ref{subsect:choice}.

Once the embedding $\iota$ of $\Cyl_S$ into the diffeomorphism invariant symmetric sector is constructed,
one can use it to induce dynamics from the full to the reduced theory.
This will allow one to understand cosmological consequences of different choices of dynamics in the full theory,
providing a key way to understand and evaluate such proposals,
as well as to understand their relation to the LQC dynamics which have been successful thus far.

\subsubsection*{The volume embedding.}

There is, in fact, a `mathematically natural' choice of embedding $\iota$ into diffeomorphism invariant states, the `volume embedding', which,
however, does not satisfy the condition (\ref{iotasymm}), and is not physically satisfactory in other ways as well.

Nevertheless, because  of its remarkable simplicity,
and because it has not yet appeared in the literature, we mention it here.
It is defined as follows.
Fix a basis $\mathcal{B}$ of $\Cyl$ consisting in spin networks $|\gamma, \vec{j}, \vec{T}\rangle$
which are eigenstates of the spatial volume.  Let $V(\gamma, \vec{j}, \vec{T})$ denote the
spatial volume eigenvalue corresponding to $|\gamma, \vec{j}, \vec{T}\rangle$. Then one can define
an embedding $\iota: \Cyl_\HIdec \rightarrow \Cyl\dual$ by
\begin{equation}
(\iota \psi | \gamma, \vec{j}, \vec{T} \rangle :=
\langle \psi | V(\gamma, \vec{j}, \vec{T}) \rangle_V
\end{equation}
for all $|\gamma, \vec{j}, \vec{T} \rangle \in \mathcal{B}$, where $|V'\rangle_V$
denotes the eigenstate of volume $\hat{V}:= \hat{p}^{3/2}$
in LQC with eigenvalue $V'$.  Because the spatial volume is
diffeomorphism invariant, the state $(\iota \psi |$ is manifestly diffeomorphism invariant,
and so one has a very simple, natural-looking embedding of LQC into diffeomorphism invariant LQG.

However, as mentioned, there is a problem with this embedding: The states in its image are very clearly not homogeneous
or isotropic.  This can be seen, for example, from the explicit expression for the embedding of an
eigenstate $\psi = |V'\rangle_V$ of volume:
\begin{equation}
\label{eqn:volembed}
(\iota V' | = \sum_{\substack{|\gamma, \vec{j}, \vec{T}\rangle \in \mathcal{B}\\
s.t. V(\gamma, \vec{j}, \vec{T}) = V' }} \langle \gamma, \vec{j}, \vec{T} | .
\end{equation}
Each such state in the image of $\iota$ is an equal superposition of all full theory
spin networks with a given total volume.  Thus, included in this equally weighted superposition
are quantum states which approximate arbitrary spatial geometries compatible with the given total
spatial volume.  In particular, it is clear that this superposition is in no way peaked on any geometries
which are homogeneous and isotropic.\footnote{The 
state (\ref{eqn:volembed}) is diffeomorphism invariant, and therefore invariant under any particular action of any spatial symmetry group. 
But this is true of \textit{any} full theory state which satisfies the diffeomorphism constraint, and hence is trivial as an additional restriction. In particular, this has nothing to do with the notions of homogeneity and isotropy assumed in cosmology. See \cite{engle2006} for a full discussion.
}
Even worse than this,
the above definition depends critically on the choice of basis $\mathcal{B}$ used to define it, a basis for which it is not
at all clear whether a natural choice exists.
It is because of these deficiencies that we advocate a more systematic approach such as that presented in this paper.

\subsection{Classical analysis and consequent clarification of the embedding strategy}
\label{sec:clarify}

\subsubsection{Operators of interest are expected to \textit{not} preserve $\Vsymm$}

In the following, we are interested in the question of whether or not a given operator $\hat{O}$ 
is expected to preserve the symmetric sector $\Vsymm$. This will be true in the quantum theory 
if the commutator of $\hat{O}$ with the symmetry constraint operators 
$\hat{\mathbb{S}}[f,g]$ are equal to
sums of compositions of operators with symmetry constraint operators on the rightmost side. This in turn will be true 
only if the classical analogue $O$ of $\hat{O}$ has Poisson bracket with the constraint functions $\mathbb{S}[f,g]$
again equal to symmetry constraint functions $\mathbb{S}[f,g]$ with $f$ and $g$ possibly phase-space dependent,
without any complex conjugate of these contraint functions appearing.

Using the Poisson brackets calculated in section \ref{sec:classical},
it is easy to see this is \textit{not} the case for at least two of the simplest quantities of interest. 
Specifically, from equations (\ref{SC:rYdef},\ref{SC:YVbra},\ref{SC:rTdef}), we find that 
\begin{align}
\{\mathbb{S}[f,g], V\}
= \frac{\newton \gamma}{2} Y[f] V[g] - \frac{\newton \gamma}{2} V[f] Y[g]
= \frac{\newton \gamma}{2} T[f,g] = \frac{\newton \gamma}{4 \alpha} 
\left( \mathbb{S}[f,g] - \overline{\mathbb{S}[f,g]}\right)
\end{align}
so that it fails for the volume of the universe.  Furthermore, as already pointed out from 
(\ref{SC:SHbra}) it fails for the Euclidean Hamiltonian constraint 
$C_{\text{E}}$. 
That is, in the full theory, we expect $\hat{V}$ and $\hat{C}_{\text{E}}$
to not preserve $\Vsymm$.
Note, this is different from the simpler Bianchi I case \cite{behm2017}, in which both the Euclidean and Lorentzian Hamiltonian
constraints, as well as the volume operator, preserve the symmetric sector.

\subsubsection{Consequent clarification of strategy}

The consequence of this is that, of the two possibilities for fixing the remaining freedom in $\iota$,
(\ref{intertwine}) and (\ref{eqn:matel}), it is necessary to use (\ref{eqn:matel}) --- equality of matrix elements.
However, from our experience with the Bianchi I case \cite{behm2017}, we {\it also}
expect the states in the symmetric sector to be non-normalizable, which makes the computation of matrix elements
in (\ref{eqn:matel}) not straightforward.  The strategy we propose is to make the states normalizable in 
the \textit{diffeomorphism-invariant inner product}, by introducing a \textit{cut off} which is then later removed.

Let us discuss a concrete proposal for such a cut-off.  First, let $\mathcal{S}$ denote the set of all 
diffeomorphism equivalence classes of spin-networks belonging to some fixed spin-network basis of $\Cyl$. 
For each $s \in \mathcal{S}$, let $\Psi_s^{\diff}$ denote the image, under the rigging map $\eta$, of the 
spin-network state associated to any representative $(\gamma, \vec{j}, \vec{T})$ in $s$. Finally, let $\psi_p$ denote 
the eigenstate of $\hat{p}$ in LQC with eigenvalue $p$. 
Suppose we then solve for the embedding $\iota$ by starting with the 
Ansatz
\begin{align*}
(\iota \psi_p | = \sum_{s \in \mathcal{S}} \lambda_{s,p} \Psi_s^{\diff} \in \Cyl^\star_{\Diff} .
\end{align*}
To make $\iota \psi_p$ normalizable in $\Hil_\diff$, it is then sufficient to simply restrict the above sum to a finite 
subset of $\mathcal{S}$. This task becomes much more 
manageable if we choose the ``extended diffeomorphisms'' \cite{fr2004} in solving the diffeomorphism constraint --- that is, in defining $\Cyl_\Diff^\star$, the rigging map $\eta$, and hence also in defining $\mathcal{S}$. 
Then $\mathcal{S}$ starts out at least countable. 
A consequence of such a choice is that the volume operator must
be quantized in the manner introduced by Rovelli and Smolin \cite{rs1994} as opposed to that introduced by Ashtekar and Lewandowski \cite{al1997}.
The task of defining a finite subset of $\mathcal{S}$ 
is further simplified by restricting consideration to the subset $\mathcal{S}_{\text{cell}}$ of $\mathcal{S}$ consisting 
in elements for which the graph is dual to a cell complex. This eliminates different possible knottings of a given abstract graph from being distinctly considered in the sum. Such an elimination involves no loss of physics due to an absence of observables  which are sensitive to distinct knottings \cite{bms2006}.
Finally, given any positive integer $N$ and positive half integer $J$, let $\mathcal{S}_{N,J}$ denote the further subset 
of $\mathcal{S}_{\text{cell}}$ consisting in elements for which the number of edges in the graph is less than $N$, and all spin labels are less than $J$.
The set $\mathcal{S}_{N,J}$ is then finite, so that 
\begin{align*}
(\iota_{N,J} \psi_p | = \sum_{s \in \mathcal{S}_{N,J}} \lambda_{s,p} \Psi_s^{\diff} \in \Cyl^\star_{\Diff} .
\end{align*}
is normalizable for each $N$, $J$, and $p$. This provides a regularized Ansatz for $\iota$, and the regulator can 
be removed by simply letting $N$ and $J$ go to infinity in an appropriate manner.

The above choice of regulator is just one possible choice, which we have laid out as an example.  
When $\iota$ is explicitly solved for, one may find that another choice of regulator is more convenient
or natural.

% We assume from now on that an option such as one of these has been chosen.

\subsection{Equivalence of projector and embedding strategies}

We close this section with a note on the equivalence of the `embedding strategy' for relating LQC and LQG presented above with the `projection strategy'  advocated in \cite{aw2009, as2011a}.
The present paper deals with the problem of relating a given quantum theory, with state space $\Hil$ and operators
$\hat{O}^i$, to a spatial symmetry reduction thereof, with state space $\Hil_S$ and corresponding
% (not necessarily independent)
operators $\hat{O}^i_S$.
In doing this, we here follow the general strategy of specifying an \textit{embedding} $\iota: \Hil_S \hookrightarrow \Hil$
of $\Hil_S$ into an appropriate `symmetric sector' defined to consist in the states of $\Hil$
satisfying operator equations expressing the relevant symmetry.
We wish to emphasize that this strategy is fully equivalent to the strategy presented
in the papers \cite{aw2009, as2011a}, in which a \textit{projection} from the larger space of states
$\Hil$ to the smaller space $\Hil_S$ is specified, the interpretation being that of `integrating out the non-symmetric degrees of freedom'.
This equivalence arises from the fact that the adjoint of
every surjective projection $\mathbb{P}: \Hil \rightarrow \Hil_S$ is injective, and hence an embedding
$\iota := \mathbb{P}^\dagger: \Hil_S \hookrightarrow \Hil$, and vice versa.
% providing a one-to-one correspondence between possible prescriptions in the two strategies.
Furthermore, $\mathbb{P}$ intertwines a pair of operators $\hat{O}, \hat{O}_S$ if and only if the corresponding
embedding $\iota = \mathbb{P}^\dagger$ also intertwines their adjoints.

Here we use the embedding perspective because it allows a clear and systematic sense in which homogeneity and isotropy
play a role. However, the work \cite{aw2009} has achieved something remarkable which will play a central role in
the companion paper \cite{behm2017}.
Specifically, the  authors have constructed a \textit{dynamical projection} from the quantum Bianchi I model to isotropic LQC,
which intertwines the \textit{Hamiltonian constraints} of the two models, whence
the corresponding embedding also intertwines the Hamiltonian constraints.  We shall call this the AW projection.
This shows that LQC passes a first test of its ability to model the dynamics of a less symmetric quantum model.
However, the role of homogeneity and isotropy implicitly used in the construction of this dynamical projector does
not have a clear generalization to the full theory.
The adjoint of this projector, however, is an \textit{embedding} of isotropic LQC into Bianchi I LQC of precisely
the type introduced above, where the role of homogeneity and isotropy is clear and has a clear
generalization to the full theory, namely that presented in this paper.  
This will be shown in detail in the companion paper \cite{behm2017}.

%On the other hand, the role of homogeneity and isotropy in defining the full theory embedding $\iota_w$ discussed here
%is clear. The aim of both the present paper and its companion \cite{behm2017} is to recast the success of \cite{aw2009} in terms of an embedding such as described in this section,  and to use this embedding to gain more insight into the Bianchi I and  isotropic LQC models, as well as to learn lessons
%relevant for constructing the corresponding embedding into the full theory.

\section{Conditions on $\iota$ to impose or test}
\label{subsect:choice}

As already noted above, in the full theory, it is expected that the intertwining condition (\ref{intertwine}) 
will not be possible to satisfy, so that equality of matrix elements (\ref{eqn:matel}) must be used instead.  
As a consequence, for the reasons given in appendix \ref{toyapp}, we first require that $\iota$ be a \textit{projective isometry}:
\begin{align}
\frac{\langle \iota \psi, \hat{O}_i \iota \phi \rangle}{\sqrt{\langle \iota \psi, \iota \psi \rangle \langle \iota \phi, \iota \phi \rangle}}
= \frac{\langle \psi, \hat{O}_i \phi \rangle}{\sqrt{\langle \psi, \psi \rangle \langle \phi, \phi \rangle}} 
\quad \text{for all }\psi, \phi \in \Hil_S .
\end{align}
The reason we require only projective isometricity and not exact isometricity is that, from the work on the embedding into Bianchi I \cite{behm2017}, we expect that $\iota$ \textit{cannot} be chosen exactly isometric. 
With projective isometricity, equality of matrix elements (\ref{eqn:matel}) becomes a strict generalization of intertwining 
(\ref{intertwine}).
Beyond this, two more conditions --- 
equation (\ref{eqn:matel}) for two choices of operators, or something equivalent in strength ---
are required to fix $\iota$ uniquely.
The possibilities for completing the choice of these two conditions are the topics of this section.

Two conditions of particular physical importance are (1.) matching of spatial curvature operators and 
(2.) matching of at least one of the forms of dynamics. 
Equality of matrix elements of average spatial curvature is important for consistency to ensure the same  
$k$-sector is represented in the reduced model as in the image of the embedding.
Matching of dynamics is important so that the embedding 
% at least formally
maps physical states into physical states.
Alternatively, one may be simply interested in investigating some other quantity, 
% (yes, because volume is in there)
in which case equality
of matrix elements for that quantity might be appropriate.
We list the possibilities in the subsections that follow.

\subsection{Equality of average spatial curvature matrix elements}
\label{subsect:spcurv}

The embedding defined as above will map the states of the LQC model under consideration into the homogeneous isotropic sector
$\Vsymm$ of loop quantum gravity. However, the LQC model considered here more specifically corresponds to
the case of zero spatial scalar curvature.
It is important to ensure that the embedding $\iota$ not only map
$\Hil_S$ into the homogeneous isotropic sector $\Vsymm$, but, more specifically, into states which correspond
to zero spatial scalar curvature in some sense.

To this end, we propose to choose one of the two
operators $\hat{O}_1$ in condition (\ref{eqn:matel})
to be the \textit{average spatial curvature
operator} defined in section \ref{subsect:curv}:
\begin{align}
\label{spintertwine}
\langle \iota \psi, \widehat{R_{\text{ave}}V^2} \iota \phi \rangle = 
\langle \psi, \widehat{R_{\text{ave}}V^2} \phi \rangle \quad \text{for all }\psi, \phi \in \Hil_S .
\end{align}
If this operator is correctly quantized on the reduced Hilbert space $\Hil_S$, it should be identically zero,
so that this condition becomes
\begin{align}
\label{spzero}
\langle \iota \psi, \widehat{R_{\text{ave}}V^2} \iota \phi \rangle = 0.
\end{align}
\textit{Side note:} If we consider embedding a cosmological model corresponding instead to positive or negative
spatial curvature ($k=1,-1$), (\ref{spzero}) no longer applies, and one must use (\ref{spintertwine}).

\subsection{Equality of volume matrix elements}
\label{subsect:vol}

The volume operator is of special interest because in LQC a certain subset of eigenvalues of the volume become 
superselected, dynamically changing the spectrum.  Furthermore, the framework of LQC is simplest in the volume 
basis, and this basis is usually used.  In the full theory, there exists standard quantizations of the volume
\cite{rs1994, al1996, al1997, gt2005}, whereas in LQC the volume operator is given by $\hat{V} = |\hat{p}|^{3/2}$.

\subsection{Equality of Hubble rate matrix elements}
\label{subsect:hubble}

At homogeneous isotropic phase space points, from (\ref{eqn:hired}) and (\ref{SC:rYdef}),
\begin{align*}
HV = \frac{\Im \mathbb{B}[\ident]}{6 \alpha \gamma} = \frac{Y[\ident]}{6\gamma},
\end{align*}
where $H$ is the Hubble rate.
This expression, when extended to generic phase space points, we interpret as an `average Hubble rate',
\begin{align*}
H_{\text{ave}}V = \frac{Y[\ident]}{6 \gamma},
\end{align*}
which is readily quantized using (\ref{SC:rYdef}):
\begin{align}
\label{qHavedef}
\widehat{H_{\text{ave}}V} := \frac{\hat{Y}[\ident]}{6\gamma} = \frac{\left[\hat{B}[\ident], \hat{V}\right]}{3i\hbar \newton \gamma^2}
\end{align}
To quantize $(H_{\text{ave}} V)$ on the LQC Hilbert space,
we note that, on the LQC phase space,
the Hamiltonian constraint is proportional to the Euclidean self-dual Hamiltonian constraint
\cite{aps2006},  which is equal to $B[\ident]$.  Specifically, 
\begin{align}
\label{BHam}
B[\ident] \approx - 2 \gamma^2 C[1]
\end{align}
where $C[1]$ is the gravitational part of the Hamiltonian constraint
using the conventions of \cite{aps2006}.  One may thus define
\begin{align}
\label{qBHamAPS}
\hat{B}[\ident]_S :=  - 2 \gamma^2 \hat{C}^{\text{APS}}_{\text{grav}}
\end{align}
where $\hat{C}^{\text{APS}}_{\text{grav}}$ is as in \cite{aps2006, mop2011}.
$\widehat{H_{\text{ave}} V}_S$ is then given by the same expression as (\ref{qHavedef}), 
with $\hat{B}[\ident]$ and $\hat{V}$ replaced by $\hat{B}[\ident]_S$ and $\hat{V}_S := \hat{p}^{3/2}$,
defining the corresponding operator in LQC.

\subsection{Equality of Hamiltonian and Master constraint matrix elements}
\label{subsect:candyn}

There are two different proposals for the definition of the canonical dynamics: The (remaining) Hamiltonian constraint 
$\hat{C}[1]$ and Master constraint $\hat{M}$.
With $\iota$ fixed, each choice of Hamiltonian constraint $\hat{C}[1]$ or master constraint $\hat{M}$
in the full theory will then induce a unique corresponding constraint in LQC via matrix elements
\begin{align*}
\langle \iota \psi, \hat{C}[1] \iota \phi \rangle &= \langle \psi, \hat{C}\phi \rangle \\
\langle \iota \psi, \hat{M} \iota \phi \rangle &= \langle \psi, \hat{M} \phi \rangle
\end{align*}
for all $\psi, \phi \in \Hil_S$.
These induced operators can then be compared
with proposals for these operators in LQC already in the literature \cite{aps2006, bt2006},
$\hat{C}_S$ and $\hat{M}_S$. Alternatively, one might try to \textit{impose} one of the above
two conditions as the second condition on $\iota$ used to \textit{define} $\iota$.
It is by no means guaranteed that this is possible, so the existence of such an $\iota$ would already
be a strong indicator of compatibility of the relevant dynamics in the full and reduced theories.

\subsection{Equality of spin foam amplitudes}
\label{subsect:sfdyn}

The other approach to dynamics in the full theory is via \textit{spin-foam models}.
The basic structure provided by a spin-foam model is the `transition amplitude function' $\eta(\Psi, \Phi)$ between diffeomorphism-invariant states $\Psi, \Phi$, related to the projector $P_{\text{ph}}$ onto physical states and the inner product thereon
by $\eta(\Psi, \Phi) = \langle P_{\text{ph}} \Psi, P_{\text{ph}} \Phi \rangle_{\text{ph}}$.
In LQC one similarly has a projector
$P^S_{\text{ph}}$ onto physical states and physical inner product $\langle , \rangle_{\text{ph}}^S$ ,
both defined through group averaging using
the (self-adjoint version of) the Hamiltonian constraint, so that one also has an LQC transition amplitude function $\eta_S (\psi, \phi)$, $\psi, \phi \in \Hil_S$ \cite{ach2009,ach2010,chn2010,hrvw2010, ach2010b}.
Given a proposal for spin-foam dynamics in the full theory, one can then use the embedding
$\iota$ to define a corresponding induced dynamics on LQC via the condition
\begin{align}
\label{sfinduce}
\eta(\iota \psi, \iota \phi) = \eta(\psi,\phi)
\end{align}
for all $\psi, \phi \in \Hil_S$.
$\eta(\cdot, \cdot)$ can then be compared with the spin-foam amplitude $\eta_S(\cdot, \cdot)$
arising from the standard dynamics for LQC used up until now \cite{bt2006, ach2009, ach2010, ach2010b}.
Alternatively, as a replacement for one of the two intertwining conditions used to define $\iota$,
one can try to \textit{impose} that these two spin-foam amplitudes in LQC be equal.
Equation (\ref{sfinduce}) would then become a condition on $\iota$,
restricting $\iota$ just as much as a condition of the form (\ref{intertwine}) or (\ref{eqn:matel}) would.

\subsection{Remark on embedding into Bianchi I}

On the Bianchi I phase space, as on the homogeneous isotropic phase space
(\ref{qBcomplexify},\ref{BHam}), the quantities
$C[1]$, $\mathbb{B}[\ident]$ and $V$
are related by $\mathbb{B}[\ident] = - 2 \gamma^2 \mathbb{C}(C[1])$
with $V$ as the complexifier.
Thus, in Bianchi I LQC, the corresponding three operators have the same relationship
to each other as in isotropic LQC (\ref{qBHamAPS}):
\begin{align*}
\hat{\mathbb{B}}[\ident] = -2 \gamma^2 e^{-\hat{V}} \widehat{C[1]} e^{\hat{V}}.
\end{align*}
Therefore, in applying the strategy of this paper to embed isotropic LQC into Bianchi I,
if any two of $\hat{\mathbb{B}}[\ident]$, $\hat{V}$, $\widehat{C[1]}$ are intertwined,
so is the third one.
That is, in the simpler case of embedding into Bianchi I LQC, the stronger, intertwining version of
the criteria in sections \ref{subsect:spcurv}, \ref{subsect:vol}, \ref{subsect:hubble},
\textit{and} \ref{subsect:candyn} above are all achieved.s
The details of this are discussed in the companion paper \cite{behm2017}.
The remarkably clean and broad success in applying the strategy of this paper to embed into Bianchi I LQC provides
hopeful confidence in completing its application to the full theory.

\section*{Acknowledgements}

%%%%%%%%%%%%%%%%%%%%%%%%%%%%%%% Begin Ack %%%%%%%%%%%%%%%%%%%%%%%%%%%%%%%%%%%%%%%%%%%%%%%%

The authors thank Ted Jacobson, Atousa Chaharsough Shirazi, Brajesh Gupt, Jorge Pullin, Parampreet Singh, Xuping Wang, and Shawn Wilder for helpful discussions.
This work was supported in part by NSF grants PHY-1205968 and PHY-1505490, and by NASA through the University
of Central Florida's NASA-Florida Space Grant Consortium.

%%%%%%%%%%%%%%%%%%%%%%%%%%%%%%% End Ack %%%%%%%%%%%%%%%%%%%%%%%%%%%%%%%%%%%%%%%%%%%%%%%%%%

\appendix

\renewcommand\thesubsection{\Alph{section}.\arabic{subsection}}
%
% So a subsection in an appendix doesn't say "appendix"

\section{Necessary and sufficient condition for maximal symmetry}
\label{app:maxsym}

This appendix will show that the Riemann curvature of a $d$-dimensional (pseudo-)Riemannian manifold $(M, g_{ab})$ has the constant-curvature form
\begin{equation}\label{Rmaxsym}
        R_{abd}{}^e = \frac{2}{d (d - 1)}\, R\, g_{d[a}\, \delta_{b]}{}^e,
\end{equation}
with the scalar curvature $R$ constant throughout $M$, if and \textit{only if} $(M, g_{ab})$ admits a space of Killing fields of the maximal dimension $d (d + 1) / 2$.  That is, the curvature of a geometry has the usual maximally symmetric form if and  only if the geometry is, in fact, maximally symmetric.  Proving that a maximal space of Killing fields implies a curvature of the form (\ref{Rmaxsym}) is standard \cite{wald1984}, so we will focus on the converse.

It is well known \cite{wald1984} that any Killing field $\xi^c$ of $(M, g_{ab})$ is uniquely determined by its Killing data
\begin{equation}
        \xi^c(p)
        \qquad\text{and}\qquad
        \Xi_d{}^e(p) := (\grad\!_d\, \xi^e)(p)
        \quad\leadsto\quad
        \Xi_{de}(p) = - \Xi_{ed}(p)
\end{equation}
at any one point $p \in M$.  One shows this explicitly by integrating the coupled system of first-order ordinary differential equations
\begin{equation}\label{Ktran}
        \mfs{D}_{\dot\gamma} \bigl( \xi^c \oplus \Xi_d{}^e \bigr)
                := \bigl( \grad\!_{\dot\gamma}\, \xi^c - \dot\gamma^a\, \Xi_a{}^c \bigr)
                        \oplus \bigl( \grad\!_{\dot\gamma}\, \Xi_d{}^e - \xi^m\, \dot\gamma^a\, R_{mad}{}^e \bigr)
                = 0
\end{equation}
along a curve $\gamma(t)$ running from $p = \gamma(0)$ to any other point $q = \gamma(1) \in M$.  The resulting Killing data at $q$ include, in particular, the value $\xi^c(q)$ of the Killing field there.  This reconstruction gives a \textit{unique} set  of Killing data at $q$, however, only if the result does not depend on the path $\gamma(t)$.  To check whether this is so, it is sufficient to introduce two vector fields $X^a$ and $Y^b$ on $M$ and calculate the commutator
\begin{equation}
        \comm{\mfs{D}_X}{\mfs{D}_Y} \bigl( \xi^c \oplus \Xi_d{}^e \bigr)
                = \mfs{D}_{\comm{X}{Y}} \bigl( \xi^c \oplus \Xi_d{}^e \bigr)
                        + \bigl( 0 \oplus (- X^a\, Y^b\, \mfs{K}_{abd}{}^e) \bigr),
\end{equation}
where we have used the Bianchi identities and defined
\begin{equation}\label{Kdef}
        \mfs{K}_{abd}{}^e := \xi^m\, \grad\!_m R_{abd}{}^e
                + \Xi_a{}^m\, R_{mbd}{}^e
                + \Xi_b{}^m\, R_{amd}{}^e
                + \Xi_d{}^m\, R_{abm}{}^e
                - \Xi_m{}^e\, R_{abd}{}^m.
\end{equation}
In mathematical terms, $\mfs{D}_X$ is a connection on the bundle of Killing data over $M$, which is the direct (Whitney) sum of the tangent bundle and the bundle of 2-forms with one index raised using the metric.  What we have calculated in the last term  here is precisely the curvature of that connection, which as always measures the obstruction to the (path-independent) integrability of the holonomies of $\mfs{D}_X$.  Importantly, given a field of Killing data $\xi^c \oplus \Xi_d{}^e$, the tensor $\mfs{K}_{abd}{}^e$  is just the Lie derivative of $R_{abd}{}^e$ along the vector field $\xi^m$, with all derivatives $\grad\!_m \xi^n$ of the vector field replaced by $\Xi_m{}^n$.  Thus, if $(M, g_{ab})$ actually admits a Killing field $\xi^c$, and we choose $\Xi_d{}^e = \grad\!_d\,  \xi^e$ to complete the field of Killing data, then $\mfs{K}_{abd}{}^e = \Lie_\xi R_{abd}{}^e = 0$, and the Killing transport defined above is path-independent.

Now suppose we know only that the Riemann curvature of $(M, g_{ab})$ satisfies (\ref{Rmaxsym}) with $R$ constant throughout $M$.  The first term in (\ref{Kdef}) then vanishes for any $\xi^m$, and the remaining terms cancel one another for any $\Xi_d{}^e  = - \Xi^e{}_d$.  It follows that the transport equations (\ref{Ktran}) will be integrable (\textit{i.e.}, path-independent) for \textit{any} choice of Killing data at an arbitrary starting point $p$.  One can therefore reconstruct a unique Killing field throughout  $M$ for any choice of the Killing data at a point, and $M$ admits a maximal space of Killing fields.

\section{A toy example demonstrating the strategy}
\label{toyapp}

\subsection{The toy model and the `quantum symmetric sector'}

Consider a phase space $\Gamma$ of $N$ degrees of freedom, with canonical coordinates
$q_1, \dots q_N$, $p_1, \dots p_N$, so that the symplectic structure takes the form
\begin{align*}
\Omega = \sum_{i=1}^{N} dp_i \wedge dq_i
\end{align*}
and one has basic Poisson brackets
%
% Recall: When \Omega = dp \wedge dq, the basic Poisson brackets are \{q,p\} =1.
% The reason for the flip in the order of q and p is due to the definition of the inverse of the Symplectic structure:
% \Omega_{\alpha\beta} \Omega^{\beta \gamma} = \delta^\gamma_\alpha.
% It is because the second index of the first factor is contracted with the first index of the second
% that the order flips.
%
\begin{align}
\label{fulltoypoisson}
\{q_i, p_j\} = \delta_{i,j} .
\end{align}
Suppose we impose the following condition of `homogeneity' in this phase space,
\begin{align}
\label{toyqpsymm}
q_i = q_j  \text{ and } p_i=p_j \text{ for all }i,j.
\end{align}
Let $\Gamma_S$ denote the submanifold of $\Gamma$ satisfying this condition.
Choose as coordinates on this submanifold
\begin{align*}
q&:= q_1 (= q_2 = \cdots = q_N) \\
p&:= p_1(=p_2= \cdots = p_N).
\end{align*}
The pull-back of $\Omega$ to $\Gamma_S$ then provides the symplectic structure on $\Gamma_S$:
\begin{align*}
\Omega_S = N dp \wedge dq
\end{align*}
yielding basic Poisson brackets
\begin{align}
\label{redtoypoisson}
\{q, p\} = 1/N .
\end{align}
Let us call $\Gamma$ the `full' theory and $\Gamma_S$ the `reduced' theory.
Schr\"odinger quantization of the Poisson algebra (\ref{fulltoypoisson}) leads to a Hilbert space of states
$\Hil$ consisting in functions of the $N$-tuple
$(q_1, \dots q_N)$, with basic operators
\begin{align*}
\hat{q}_i \Psi(q_1, \dots q_N) &:= q_i \Psi(q_1, \dots q_N) \\
\hat{p}_i \Psi(q_1, \dots q_N) &:= -i \frac{\partial}{\partial q_i} \Psi(q_1, \dots q_N).
\end{align*}
Quantization of the Poisson algebra (\ref{redtoypoisson}) leads to a space $\Hil_S$ of states
consisiting in functions of $q$ with basic operators
\begin{align*}
\hat{q} \,  \psi(q) &:= q \psi(q) \\
\hat{p} \,  \psi(q) &:= -\frac{i}{N}\frac{d}{dq} \psi(q).
\end{align*}
To quantize the `symmetry conditions' (\ref{toyqpsymm}), one needs to reformulate them so that they form a first class set.
This can be done by defining complex quantities analogous to the quantities $\mathbb{B}[f]$ in the main text:
\begin{align*}
\mathbb{B}_i := q_i + i p_i
\end{align*}
so that the symmetry conditions become
\begin{align}
\label{firstclasscond}
\mathbb{S}_{ij}:= \mathbb{B}_i - \mathbb{B}_j = 0 \quad \text{for all }i, j,
\end{align}
a manifestly first class set.  The $\mathbb{B}_i$ are represented on $\Hil$ by the operators
\begin{align}
\label{bquant}
\hat{\mathbb{B}}_i := \hat{q}_i + i \hat{p}_i =
q_i + \frac{\partial}{\partial q_i}.
\end{align}
The quantum symmetric sector within $\Hil$ is then defined as the set of states $\Psi \in \Hil$ satisfying
\begin{align}
\label{toyqsymm}
\hat{\mathbb{S}}_{ij}\Psi := \left(\hat{\mathbb{B}}_i - \hat{\mathbb{B}}_j\right) \Psi = 0 \quad \text{for all }i,j .
\end{align}
We wish to find an embedding $\iota : \Hil_S \rightarrow \Hil$ into the above symmetric sector,
that is, an embedding satisfying
\begin{align}
\label{toyiotasymm}
\hat{\mathbb{S}}_{ij} \circ \iota = 0 \quad \text{for all }i,j .
\end{align}
To solve the condition (\ref{toyiotasymm}) explicitly, one can represent $\iota$ through its
integral kernel:
\begin{align}
\label{iotakernel}
(\iota \psi)(q_1 , \dots q_N)
= \int_{\R} \iota(q_1, \dots, q_N; q) \psi(q) dq.
\end{align}
Condition (\ref{toyiotasymm}) then implies exactly that
\begin{align}
\label{iotaform}
\iota(q_1, \dots, q_N; q) =  g(q_{\text{ave}}, q) e^{-\frac{1}{2}\sum_i q_i^2}
\end{align}
for some undetermined function $g: \R \times \R \rightarrow \C$,
where $q_{\text{ave}}:= \frac{1}{N} \sum_i q_i$.
%
%*Derivation:* (keep as comment so we don't have to rederive it)
%\begin{eqnarray*}
%\left(q_1+ \frac{\partial}{\partial q_1}\right) \iota & = \cdots =  \left(q_N+ \frac{\partial}{\partial q_N}\right) \iota
%&= \left(q - \frac{1}{N} \frac{\partial}{\partial q}\right) \iota \\
%e^{-\frac{1}{2}q_1^2}\frac{\partial}{\partial q_1}\left(e^{\frac{1}{2}q_1^2} \iota\right)
%& = \cdots = e^{-\frac{1}{2}q_N^2}\frac{\partial}{\partial q_N}\left(e^{\frac{1}{2}q_N^2} \iota\right)
%&= - \frac{1}{N} e^{\frac{N}{2}q^2}\frac{\partial}{\partial q}\left(e^{-\frac{N}{2}q^2} \iota\right)\\
%\frac{\partial}{\partial q_1}\left(e^{\frac{1}{2}(\sum_i q_i^2 - N q^2)} \iota\right)
%& = \cdots =\frac{\partial}{\partial q_N}\left(e^{\frac{1}{2}(\sum_i q_i^2 - N q^2)} \iota\right)
%&= - \frac{1}{N} \frac{\partial}{\partial q}\left(e^{\frac{1}{2}(\sum_i q_i^2 - N q^2)} \iota\right)
%\end{eqnarray*}
%This last equation tells us the gradient of the expression in parenthesis, which tells us exactly that
%\begin{align*}
%e^{\frac{1}{2}(\sum_i q_i^2 - N q^2)} \iota = g(\sum_i q_i - N q)
%\end{align*}
%which yields the form quoted above.
%

\subsection{Fixing $g$ through intertwining}

One thus sees that,
after imposing that $\iota$ map into the symmetric sector,
there remains precisely the freedom to choose a function of two variables.
As we shall see, this freedom can be used to impose that $\iota$ intertwine two more operators $\hat{O}_i$,
which can be operators of interest
which one wishes to study:
\begin{align}
\label{toyintertwine}
\hat{O}_i \circ \iota = \iota \circ \hat{O}_i .
\end{align}
Specifically, for the purpose of this example, we choose these two operators to be
\begin{align*}
\hat{q}_{\text{ave}}:= \frac{1}{N} \sum_{i=1}^N \hat{q}_i,
\qquad
\hat{p}_{\text{ave}}:= \frac{1}{N} \sum_{i=1}^N \hat{p}_i.
\end{align*}
The corresponding operators on $\Hil_S$ are just $\hat{q}$ and $\hat{p}$.
Imposing $\hat{q}_{\text{ave}} \circ \iota = \iota \circ \hat{q}$
and $\hat{p}_{\text{ave}} \circ \iota = \iota \circ \hat{p}$ then uniquely determines $g$ to be
\begin{align*}
g(q_{\text{ave}}, q) = C \delta(q_{\text{ave}}-q)e^{\frac{1}{2}N q^2}
\end{align*}
where $C$ is an undetermined integration constant.
The resulting $\iota$ is then given by
\begin{align}
\label{eqn:resiota}
(\iota \psi)(q_1, \dots , q_N)
= C \int_{\R} \delta\left(q_{\rm ave} - q\right) e^{\frac{1}{2}\left(N q^2 - \sum_i q_i^2\right)} \psi(q) dq
=  C e^{\frac{1}{2}\left(N q_{\text{ave}}^2 - \sum_i q_i^2\right)} \psi(q_{\text{ave}}) .
\end{align}
By construction $\iota$ intertwines $\hat{q}_{\text{ave}}$ and $\hat{p}_{\text{ave}}$.  As a consequence,
$\iota$ also intertwines the following further operators
\begin{description}
\item[$\hat{\mathbb{B}}_{\text{ave}}$ and $\hat{\mathbb{B}}_{\text{ave}}^\dagger$:]
Consider
\begin{align}
\hat{\mathbb{B}}_{\text{ave}}:= \frac{1}{N}\sum_i \hat{\mathbb{B}}_i .
\end{align}
The action of the $\hat{\mathbb{B}}_i$ on $\Hil$ is given in (\ref{bquant}),
while their action on $\Hil_S$ is given by
\begin{align*}
\hat{\mathbb{B}}_i := \widehat{q_i + i p_i} = \widehat{q + ip}
%=  \widehat{q} + i \widehat{p}
= q + \frac{1}{N} \frac{d}{d q} =: \hat{\mathbb{B}}.
\end{align*}
The intertwining of $\hat{\mathbb{B}}_{\text{ave}}$ and $\hat{\mathbb{B}}_{\text{ave}}^\dagger$ then follows
from $\hat{\mathbb{B}}_{\text{ave}} = \hat{q}_{\text{ave}} + i \hat{p}_{\text{ave}}$
and $\hat{\mathbb{B}}_{\text{ave}}^\dagger = \hat{q}_{\text{ave}} - i \hat{p}_{\text{ave}}$.

\item[All of the $\hat{\mathbb{B}}_i$'s:]
Note that
\begin{align}
\hat{\mathbb{B}}_i = \frac{1}{N} \sum_j \hat{\mathbb{S}}_{ij} + \hat{\mathbb{B}}_{\text{ave}}.
\end{align}
Equation (\ref{toyiotasymm}) and the intertwining of $\hat{\mathbb{B}}_{\text{ave}}$ then implies
$\hat{\mathbb{B}}_i \circ \iota = \iota \circ \hat{\mathbb{B}}$.

\item[Simple Harmonic Oscillator Hamiltonian:]
Consider the Hamiltonian
\begin{align*}
\hat{H} = \frac{1}{2}\sum_{i=1}^N (\hat{q}_i^2 + \hat{p}_i^2)
= \frac{1}{2}\sum_{i=0}^N (\hat{\mathbb{B}}_i^\dagger \hat{\mathbb{B}}_i + 1).
\end{align*}
The corresponding operator on $\Hil_S$ is
\begin{align*}
\hat{H}_S = \frac{1}{2}\sum_{i=1}^N(\hat{q}^2 + \hat{p}^2) = \frac{N}{2} (\hat{q}^2 + \hat{p}^2)
= \frac{N}{2}(\hat{\mathbb{B}}^\dagger \hat{\mathbb{B}} + 1) .
\end{align*}
Using both of the foregoing intertwining results, we have
\begin{align*}
\hat{H} \circ \iota
& = \frac{1}{2}\sum_{i=0}^N (\hat{\mathbb{B}}_i^\dagger \hat{\mathbb{B}}_i \circ \iota + \iota)
= \frac{1}{2}\sum_{i=0}^N (\hat{\mathbb{B}}_i^\dagger \circ \iota \circ \hat{\mathbb{B}} + \iota) \\
&= \frac{N}{2}(\hat{\mathbb{B}}_{\text{ave}}^\dagger \circ \iota \circ \hat{\mathbb{B}} + \iota)
= \frac{N}{2}(\iota \circ \hat{\mathbb{B}}^\dagger \hat{\mathbb{B}} + \iota)
= \iota \hat{H}_S.
\end{align*}

\end{description}

Upon reflection, it is not hard to see why, after imposing that $\iota$ map into the symmetric sector,
one expects to retain the freedom to intertwine exactly two more operators, such as $\hat{q}_{\text{ave}}$
and $\hat{p}_{\text{ave}}$ considered above.  For, the remaining freedom is
precisely that of how to map the symmetric model $\Hil_S$ into the symmetric sector.  But if the symmetric sector
is isomorphic to the symmetric model, then this remaining freedom is equivalent to the choice of an invertible
operator  on the symmetric model. But, as $\Hil_S$ has one degree of freedom, the integral kernel
representing such an operator will have two arguments, so that it will generally be completely
fixed by two conditions. Likewise, if the symmetric model had had $M$
degrees of freedom, the integral kernel of this operator would have had $2M$ arguments, so that in general $2M$ conditions
would be required to fix the remaining ambiguity in the embedding.  $2M$ in this latter case is
nothing other than the dimension of the reduced phase space $\Gamma_S$.  Thus, in the above case,
where $M=1$, we see that the reason why two further operators can be intertwined can roughly be understood
as due to the fact that the dimension of $\Gamma_S$ is two.

%Schur's lemma: If an operator commutes with all generators in an irrep, that operator is zero.

\subsection{Isometricity and reformulation in terms of matrix elements}

We end this appendix by noting that the specific embedding $\iota$ solved for in (\ref{eqn:resiota}),
for a specific choice of $C$,
is isometric.  That is, it satisfies
\begin{align}
\label{eqn:isometric}
\langle \iota \psi, \iota \phi \rangle = \langle \psi, \phi \rangle,
\end{align}
where the inner product on the left hand side and right hand side are respectively those belonging to 
$\Hil$ and $\Hil_S$.

To see this, from (\ref{eqn:resiota}), we have
\begin{align}
\label{eqn:iotaprod}
\langle \iota \psi, \iota \phi \rangle
= \int \overline{(\iota \psi)(\vec{q})}
(\iota \phi)(\vec{q}) d^N q
= |C|^2 \int e^{(Nq_{\rm ave}^2 - \sum_{i=1}^N q_i^2)}\overline{\psi(q_{\rm ave})} \phi(q_{\rm ave}) d^N q.
\end{align}
Where we use $\vec{q}$ to denote $(q_1, \dots q_N)$.
Let $\vec{u}_1:= \frac{1}{\sqrt{N}}(1,\dots 1)$ so that $\vec{u}_1$ is a unit $N$-vector, and
\begin{align*}
\vec{u}_1 \cdot \vec{q} = \sqrt{N} q_{\rm ave}.
\end{align*}
Complete $\vec{u}_1$ into an orthonormal basis $(\vec{u}_1, \dots \vec{u}_N)$ of $\R^N$.
Define the new $N$ coordinates
\begin{align*}
\alpha_i := \vec{u}_i \cdot \vec{q},
\end{align*}
so that $\alpha_1 = \sqrt{N} q_{\rm ave}$, and 
\begin{align*}
\sum_{i=1}^N q_i^2 = \vec{q} \cdot \vec{q} = 
\vec{q} \cdot \left( \sum_{i=1}^N \vec{u}_i \vec{u}_i \cdot \vec{q} \right)
= \sum_{i=1}^N (\vec{u}_i \cdot \vec{q})^2 
= \sum_{i=1}^N \alpha_i^2.
\end{align*}
Furthermore, because $\vec{u}_i$ is orthonormal, $d^N \alpha = d^N q$.
Upon changing variables from $q_1, \dots q_N$ to $\alpha_1, \dots \alpha_N$,
the integral (\ref{eqn:iotaprod}) thus becomes
\begin{align*}
\langle \iota \psi, \iota \phi \rangle
&= |C|^2 \int e^{ - \sum_{i=2}^N \alpha_i^2}
\overline{\psi\left(N^{-\frac{1}{2}}\alpha_1\right)} \phi\left(N^{-\frac{1}{2}}\alpha_1\right) d^N \alpha
= |C|^2 \pi^{\frac{N-1}{2}}\int
\overline{\psi\left(N^{-\frac{1}{2}}\alpha_1\right)} \phi\left(N^{-\frac{1}{2}}\alpha_1\right) d \alpha_1  \\
&= |C|^2 \pi^{\frac{N-1}{2}} N^{\frac{1}{2}}\int
\overline{\psi(q)} \phi(q) d q
= |C|^2 \pi^{\frac{N-1}{2}} N^{\frac{1}{2}} \langle \psi, \phi \rangle
\end{align*}
so that (\ref{eqn:isometric}) is satisfied for $C = \pi^{\frac{1-N}{4}} N^{-\frac{1}{4}}$.

More generally, $\iota$, for \textit{any} value of $C$, is a \textit{projective isometry}:
\begin{align}
\label{projisom}
\frac{\langle \iota \psi, \iota \phi \rangle}{\sqrt{\langle \iota \psi, \iota \psi \rangle \langle \iota \phi, \iota \phi \rangle}}
= \frac{\langle \psi, \phi \rangle}{\sqrt{\langle \psi, \psi \rangle \langle \phi, \phi \rangle}} .
\end{align}
An important consequence of projective isometricity is that
the intertwining conditions (\ref{toyintertwine}) become equivalent to equality of matrix elements
\begin{align}
\label{eqn:matrixeq}
\frac{\langle \iota \psi, \hat{O}_i \iota \phi \rangle}{\sqrt{\langle \iota \psi, \iota \psi \rangle \langle \iota \phi, \iota \phi \rangle}}
= \frac{\langle \psi, \hat{O}_i \phi \rangle}{\sqrt{\langle \psi, \psi \rangle \langle \phi, \phi \rangle}} 
\quad \text{for all }\psi, \phi \in \Hil_S .
\end{align}
More specifically, one can show that imposing projective isometricity, and equality of matrix elements
(\ref{eqn:matrixeq}) for the choice of operators $\hat{q}_{\rm ave}$, $\hat{p}_{\rm ave}$ above,
again the same embedding is uniquely determined.
The resulting embedding then also yields equality of matrix elements
for $\hat{\mathbb{B}}_{\rm ave}$, $\hat{\mathbb{B}}_{\rm ave}^\dagger$, and
the simple harmonic oscillator Hamiltonian as well.
In the application to Bianchi I \cite{behm2017}, operators are found which preserve the symmetric
sector, so one can use the intertwining strategy of the last subsection.
As discussed in the main text, however, in the full theory, we do not expect 
any of the operators of interest to preserve the symmetric sector,
so that the strategy using equality of matrix elements (\ref{eqn:matrixeq}) and projective isometricity (\ref{projisom})
must be used.

%{\bf
%Two more things to try to do *if* we have time (ask Phillip and Matt to do it if the don't have other writing related tasks).
%\begin{enumerate}
%\item Can we show that $\ker \{\widehat{\textbf{b}}_i - \widehat{\textbf{b}}_j\} = \Im \iota$?
%Chris almost finished the proof at the meeting on 8/20.
%
%\item Can we find an example of a full-theory Hamiltonian which (1.) weakly Poisson commutes with the symmetry constraint equations, and (2.) has products such as $q_1 p_2$ which make it not a priori obvious how to construct the corresponding reduced  theory operator, due to operator ordering ambiguities present in the reduced theory that are not in the full theory. That way, we can have an example where our method is used to *derive* a reduced theory dynamics which is not a priori obvious.
%
%\end{enumerate}
%}

%In order to demonstrate the non-triviality of the condition on the choice of reduced theory operator for there to
%\textit{exist} an embedding intertwining the operators, we provide an example demonstrating what happens when
%one tries to intertwine the simple harmonic oscillator Hamiltonian with and operator on $\Hil_S$ which is
%simply physically different: $H_S' := ...$.  In fact, for this case one can show that there exists no $\iota$
%such that $H \circ \iota = \iota \circ H_S'$, so that the existence of such an $\iota$ provides a non-trivial
%condition of compatibility on the choice of Hamiltonian in the full and reduced theories.
%% "In fact quite non trivial"?

\vspace{1em}

%
%\bibliography{..\bibtex_files\jonsbib}
%%\bibliography{jonsbib}{}
%\bibliographystyle{ieeetr}
%% Reason for ieeetr: (1.) bibliography is in order of citation,
%% (2.) titles are given for articles, but (3.) no full author first names
%% are given, only initials, and (4.) format for authors is same for
%% both articles and books
%%
%\end{document}
%

\end{document}